\documentclass[aps,prl,twocolumn,superscriptaddress,nofootinbib]{revtex4}
\usepackage{axodraw}
\usepackage{graphicx}
\usepackage{subeqn}

\newcommand{\be}{\begin{eqnarray}}
\newcommand{\ee}{\end{eqnarray}}

\newcommand{\pp}{\pi^+ \pi^-}

\newcommand{\promille}{%
  \relax\ifmmode\promillezeichen
  \else\leavevmode\(\mathsurround=0pt\promillezeichen\)\fi}
\newcommand{\promillezeichen}{%
  \kern-.05em%
  \raise.5ex\hbox{\the\scriptfont0 0}%
  \kern-.15em/\kern-.15em%
  \lower.25ex\hbox{\the\scriptfont0 00}}
\newcommand{\la}{\langle}
\newcommand{\ra}{\rangle}

\begin{document}

%
\title{Hard Exclusive Electroproduction of $\pp$ Pairs}
%

\def\groupalberta{\affiliation{Department of Physics, University of Alberta, Edmonton, Alberta T6G 2J1, Canada}}
\def\groupargonne{\affiliation{Physics Division, Argonne National Laboratory, Argonne, Illinois 60439-4843, USA}}
\def\groupbari{\affiliation{Istituto Nazionale di Fisica Nucleare, Sezione di Bari, 70124 Bari, Italy}}
\def\groupbeijing{\affiliation{School of Physics, Peking University, Beijing 100871, China}}
\def\groupchina{\affiliation{Department of Modern Physics, University of Science and Technology of China, Hefei, Anhui 230026, China}}
\def\groupcolorado{\affiliation{Nuclear Physics Laboratory, University of Colorado, Boulder, Colorado 80309-0446, USA}}
\def\groupdesy{\affiliation{DESY, Deutsches Elektronen-Synchrotron, 22603 Hamburg, Germany}}
\def\groupzeuthen{\affiliation{DESY Zeuthen, 15738 Zeuthen, Germany}}
\def\groupdubna{\affiliation{Joint Institute for Nuclear Research, 141980 Dubna, Russia}}
\def\grouperlangen{\affiliation{Physikalisches Institut, Universit\"at Erlangen-N\"urnberg, 91058 Erlangen, Germany}}
\def\groupferrara{\affiliation{Istituto Nazionale di Fisica Nucleare, Sezione di Ferrara and Dipartimento di Fisica, Universit\`a di Ferrara, 44100 Ferrara, Italy}}
\def\groupfrascati{\affiliation{Istituto Nazionale di Fisica Nucleare, Laboratori Nazionali di Frascati, 00044 Frascati, Italy}}
\def\groupgent{\affiliation{Department of Subatomic and Radiation Physics, University of Gent, 9000 Gent, Belgium}}
\def\groupgiessen{\affiliation{Physikalisches Institut, Universit\"at Gie{\ss}en, 35392 Gie{\ss}en, Germany}}
\def\groupglasgow{\affiliation{Department of Physics and Astronomy, University of Glasgow, Glasgow G12 8QQ, United Kingdom}}
\def\groupillinois{\affiliation{Department of Physics, University of Illinois, Urbana, Illinois 61801-3080, USA}}
\def\groupmit{\affiliation{Laboratory for Nuclear Science, Massachusetts Institute of Technology, Cambridge, Massachusetts 02139, USA}}
\def\groupmichigan{\affiliation{Randall Laboratory of Physics, University of Michigan, Ann Arbor, Michigan 48109-1120, USA }}
\def\groupmoscow{\affiliation{Lebedev Physical Institute, 117924 Moscow, Russia}}
\def\groupmunich{\affiliation{Sektion Physik, Universit\"at M\"unchen, 85748 Garching, Germany}}
\def\groupnikhef{\affiliation{Nationaal Instituut voor Kernfysica en Hoge-Energiefysica (NIKHEF), 1009 DB Amsterdam, The Netherlands}}
\def\groupstpetersburg{\affiliation{Petersburg Nuclear Physics Institute, St. Petersburg, Gatchina, 188350 Russia}}
\def\groupprotvino{\affiliation{Institute for High Energy Physics, Protvino, Moscow region, 142281 Russia}}
\def\groupregensburg{\affiliation{Institut f\"ur Theoretische Physik, Universit\"at Regensburg, 93040 Regensburg, Germany}}
\def\grouprome{\affiliation{Istituto Nazionale di Fisica Nucleare, Sezione Roma 1, Gruppo Sanit\`a and Physics Laboratory, Istituto Superiore di Sanit\`a, 00161 Roma, Italy}}
\def\groupsimonfraser{\affiliation{Department of Physics, Simon Fraser University, Burnaby, British Columbia V5A 1S6, Canada}}
\def\grouptriumf{\affiliation{TRIUMF, Vancouver, British Columbia V6T 2A3, Canada}}
\def\grouptokyo{\affiliation{Department of Physics, Tokyo Institute of Technology, Tokyo 152, Japan}}
\def\groupamsterdam{\affiliation{Department of Physics and Astronomy, Vrije Universiteit, 1081 HV Amsterdam, The Netherlands}}
\def\groupwarsaw{\affiliation{Andrzej Soltan Institute for Nuclear Studies, 00-689 Warsaw, Poland}}
\def\groupyerevan{\affiliation{Yerevan Physics Institute, 375036 Yerevan, Armenia}}
\def\groupkieth{\affiliation{Permanent Address: College of William \& Mary,Williamsburg, VA 23187}}

\groupalberta
\groupargonne
\groupbari
\groupbeijing
\groupchina
\groupcolorado
\groupdesy
\groupzeuthen
\groupdubna
\grouperlangen
\groupferrara
\groupfrascati
\groupgent
\groupgiessen
\groupglasgow
\groupillinois
\groupmit
\groupmichigan
\groupmoscow
\groupmunich
\groupnikhef
\groupstpetersburg
\groupprotvino
\groupregensburg
\grouprome
\groupsimonfraser
\grouptriumf
\grouptokyo
\groupamsterdam
\groupwarsaw
\groupyerevan


\author{A.~Airapetian}  \groupmichigan
\author{N.~Akopov}  \groupyerevan
\author{Z.~Akopov}  \groupyerevan
\author{M.~Amarian}  \groupzeuthen \groupyerevan
\author{V.V.~Ammosov}  \groupprotvino
\author{A.~Andrus}  \groupillinois
\author{E.C.~Aschenauer}  \groupzeuthen
\author{W.~Augustyniak}  \groupwarsaw
\author{R.~Avakian}  \groupyerevan
\author{A.~Avetissian}  \groupyerevan
% ...

\collaboration{The HERMES Collaboration} \noaffiliation 

\author{Submitted to Physics Letters B}

%
\begin{abstract}
Hard exclusive electroproduction of $\pi^+\pi^-$ 
pairs off hydrogen and deuterium targets 
has been studied by the HERMES experiment 
at DESY.
Legendre moments $\langle P_1 \rangle$ and 
$\langle P_3 \rangle$ of the angular distributions 
of $\pi^+$ mesons in the center-of-mass frame of the pair 
have been measured for the first time.  
Their dependence on the $\pi^+\pi^-$ invariant mass  
can be understood as being due to the interference 
between relative $P$-wave (isovector) 
and $S,D$-wave (isoscalar) states of the two pions. 
The increase in magnitude of $\langle P_1 \rangle$ 
as Bjorken $x$ increases is interpreted in the 
framework of Generalized Parton Distributions as an enhancement 
of flavour \mbox{non-singlet} $q\bar{q}$ exchange for larger 
values of $x$, which leads to a sizable admixture 
of isoscalar and isovector pion pairs.
In addition, the interference between 
\mbox{$P$-wave} and \mbox{$D$-wave} states 
separately for transverse and longitudinal 
pion pairs has been studied. 
The data indicate that in the $f_2(1270)$ region at   
\mbox{$\la Q^2 \ra = 3 $ GeV$^2$} \mbox{higher-twist} 
effects can be as large as the leading-twist longitudinal 
component.

\hspace{2cm}
PACS numbers: \mbox{13.60.Le}, \mbox{14.40.Cs}, \mbox{25.30.Rw}, 
\mbox{13.88.+e}
\end{abstract}
\maketitle
%
\begin{figure*}[htb!]
 \begin{center}
\includegraphics[height=7.0cm,width=4.4cm]{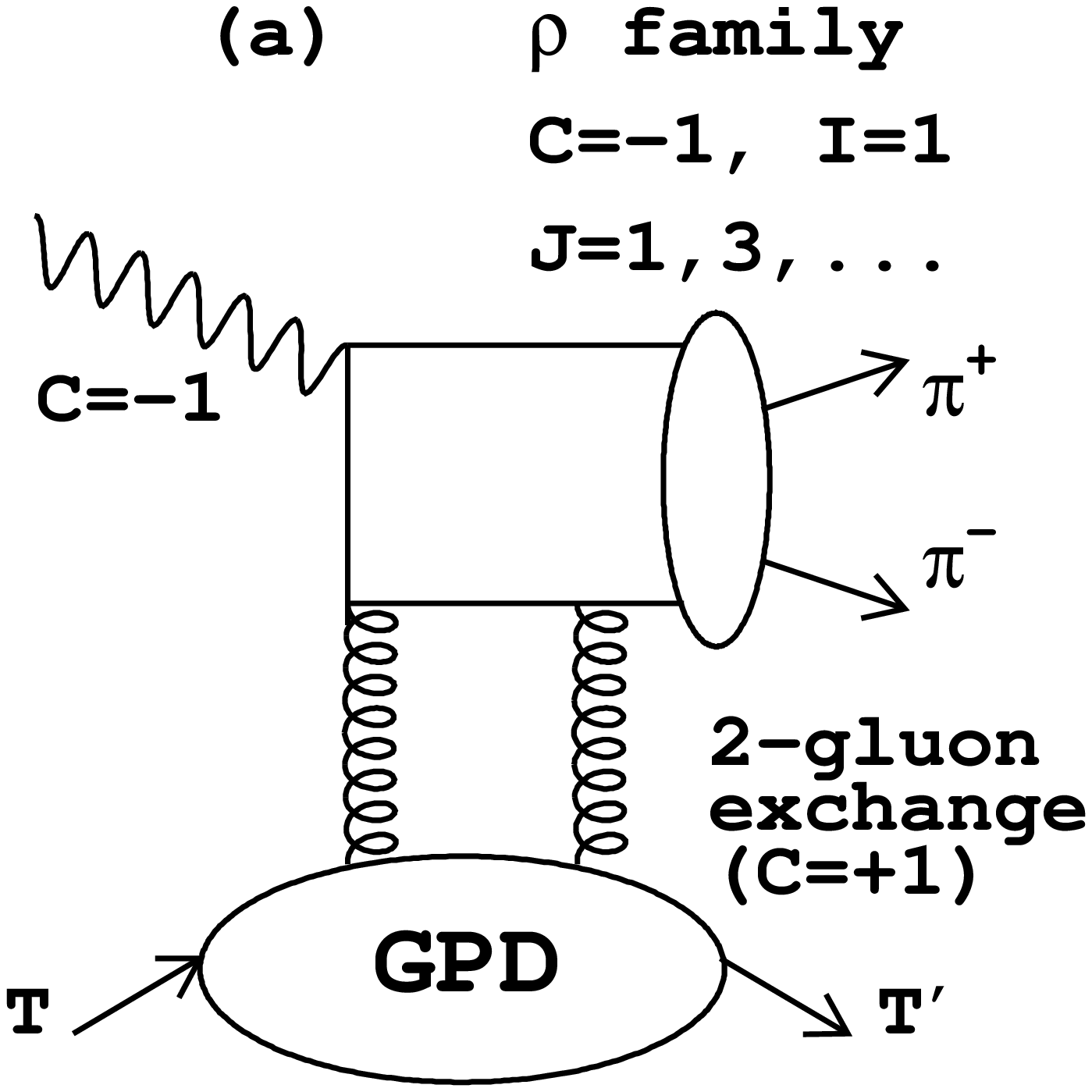}
\includegraphics[height=7.0cm,width=4.4cm]{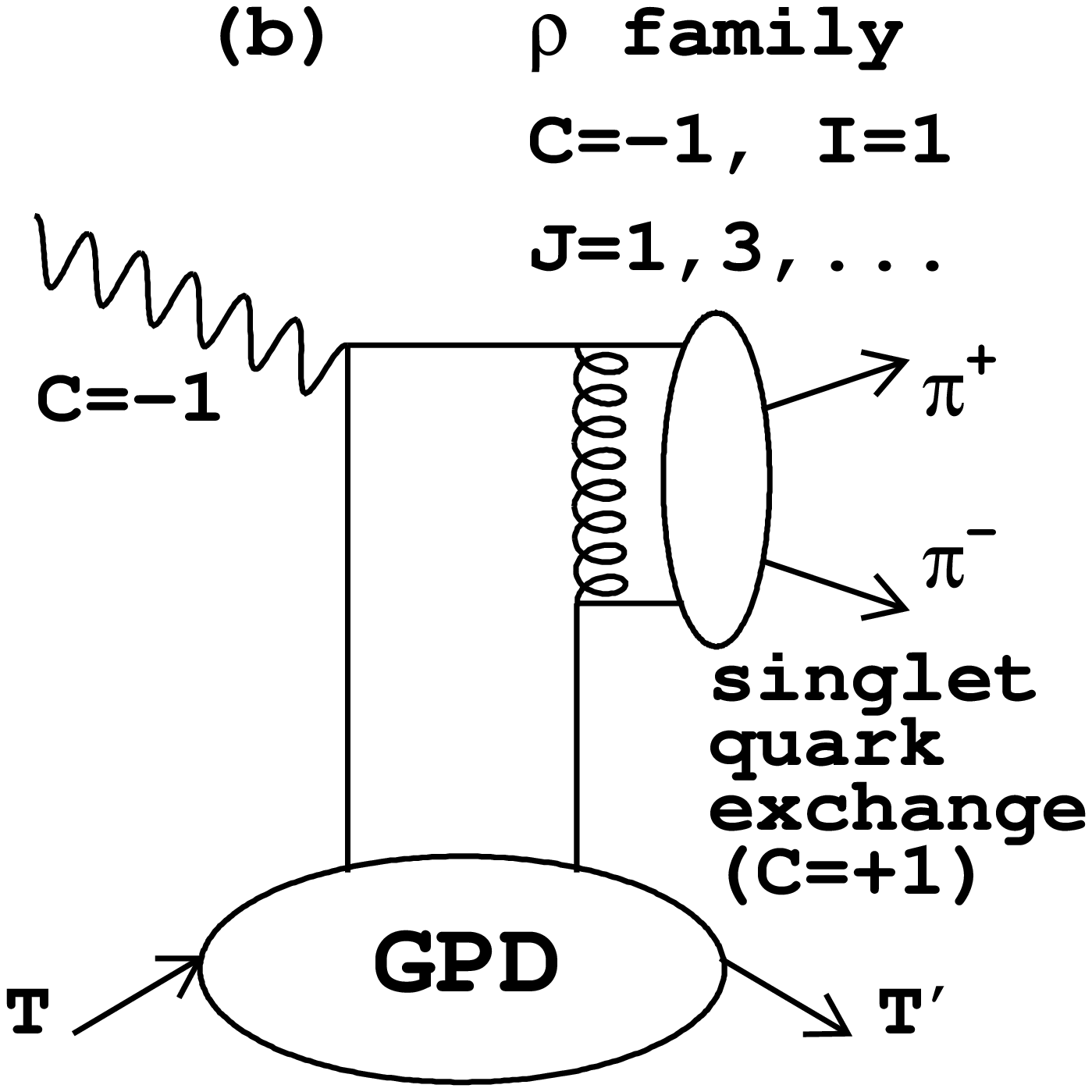}
\includegraphics[height=7.0cm,width=4.4cm]{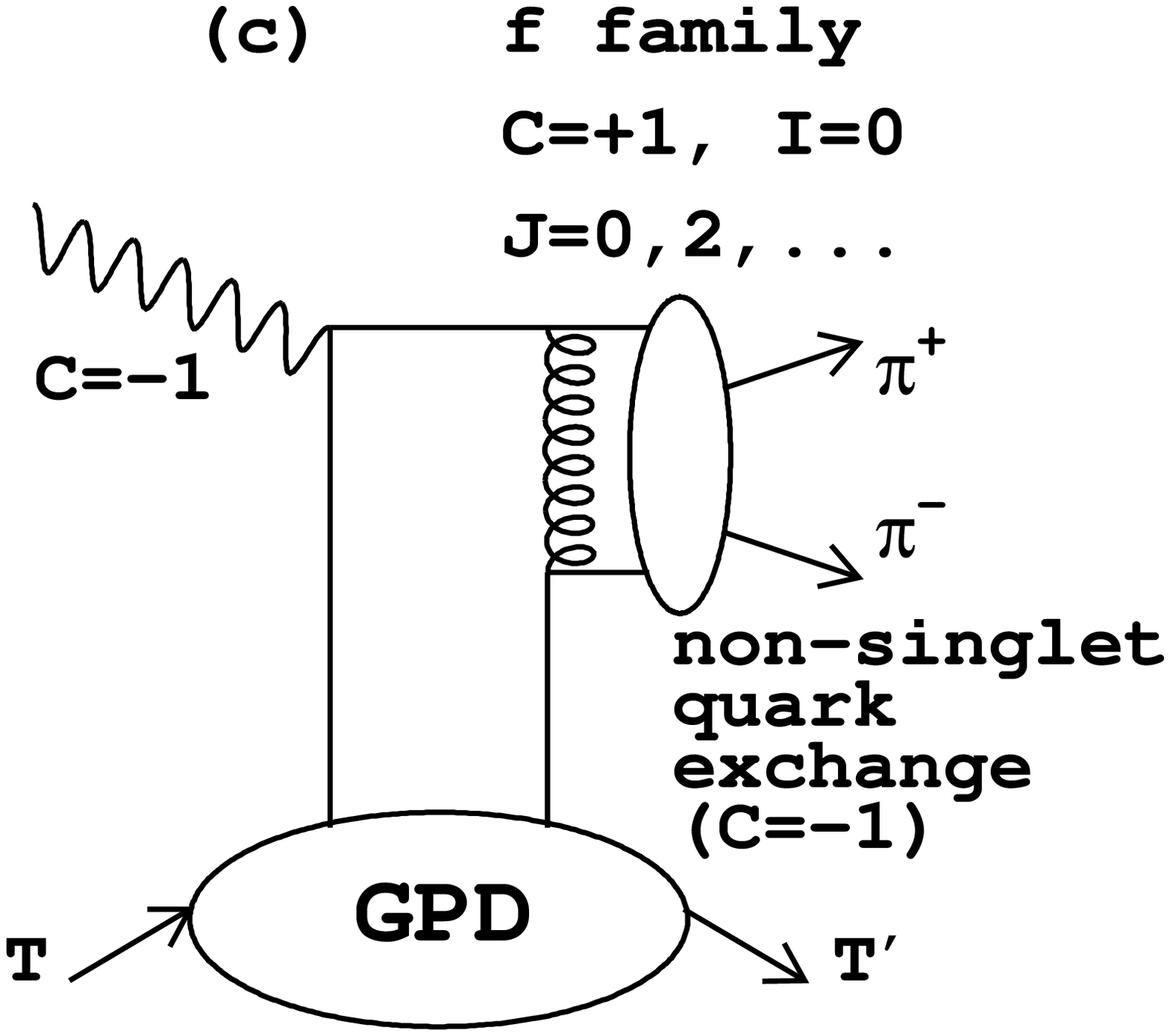}
\includegraphics[height=7.0cm,width=4.4cm]{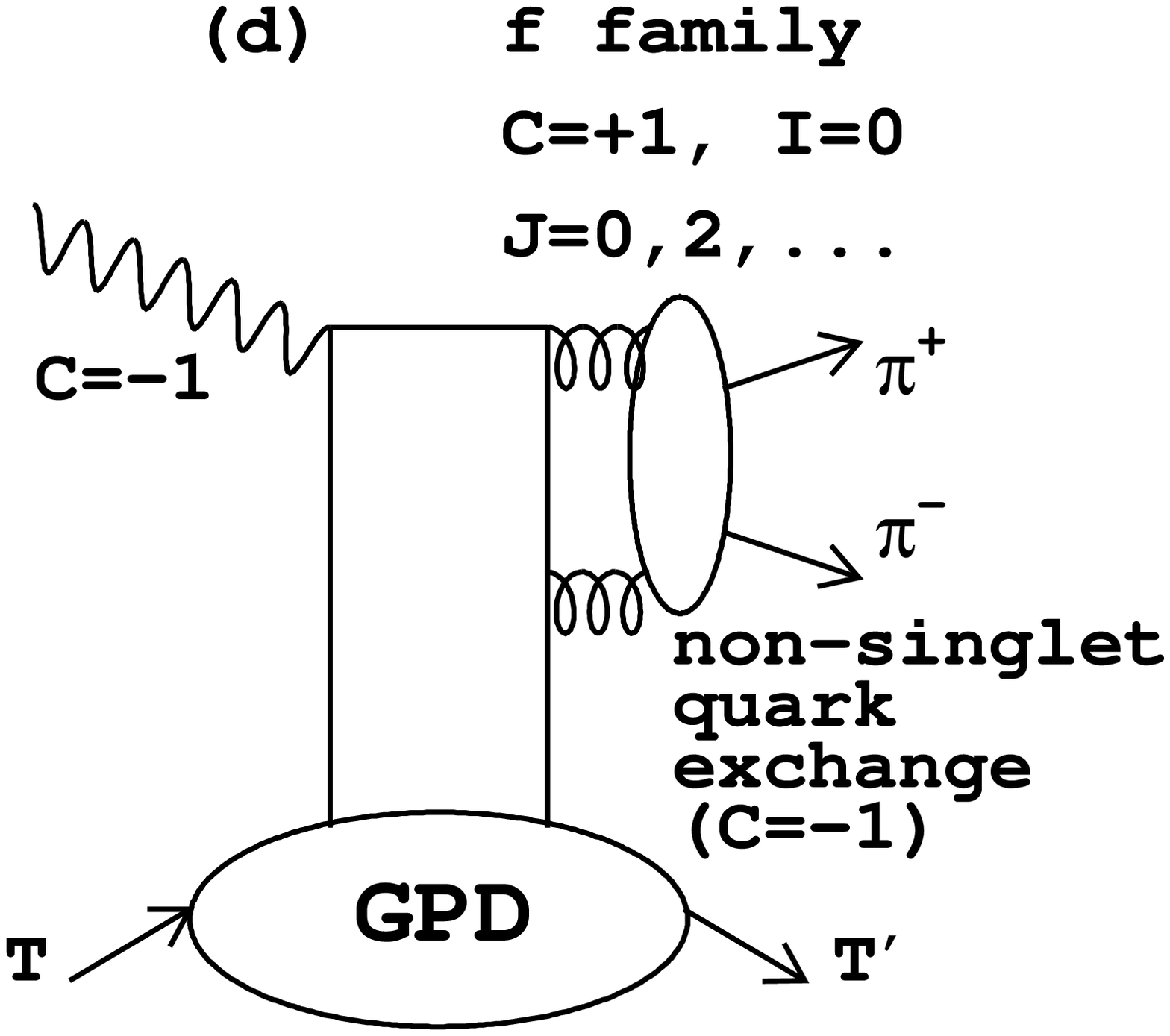}
   \caption{Leading twist diagrams for the hard exclusive
           reaction $e^+ T \rightarrow e^+ T' \; \pi^+ \pi^-$.
           Gluon exchange ({\it a)} gives rise to pions
           in the isovector state only, while
           the quark exchange mechanism ({\it b,c,d}) gives rise to
           pions in both isoscalar and isovector states.}
   \label{fig:feynman}
 \end{center}
\end{figure*}

Much of our current knowledge of the 
quark-gluon structure of the nucleon 
comes from inclusive and \mbox{semi-inclusive} 
deep inelastic scattering experiments, 
from which parton distribution functions 
can be extracted. 
However, our understanding of quark-gluon 
dynamics can be extended considerably 
by measurements sensitive to the 
generalized parton distributions 
\mbox{(GPD)~\cite{muller,radyushkin,ji98}}, 
which also describe  
the dynamical correlations between 
partons with different momenta. 
Experimentally, GPDs can be investigated 
through the analysis of hard exclusive processes 
such as the production of mesons by longitudinal  
virtual photons. 
Under these conditions the amplitude
factorizes into a hard scattering term governed by 
perturbative QCD and two soft parts, 
the GPDs for the nucleon and the distribution 
amplitude for meson formation~\cite{collins,freund}.
Hard exclusive electroproduction of 
$\pp$ pairs is sensitive to the 
interference between isospin $I=1$ and $I=0$ 
channels, and provides a new constraint on certain 
combinations of GPDs. 

This letter reports the first experimental data for  
hard exclusive $\pp$ pair production 
\begin{eqnarray}
   e^+ p \to e^+ p \ \pp  &
   \mathrm{and}  &
   \ e^+ d \to e^+ d \ \pp  \ . 
\end{eqnarray}
For the proton target, the results are interpreted 
in the GPD framework by comparing with 
predictions~\cite{polyakov01,lehmann,polyakov00}, 
thus providing valuable information for further 
modelling of GPDs.
So far, predictions exist only for the proton target.  
Exclusive pair production includes contributions from 
both \mbox{two-gluon} and \mbox{quark-antiquark} 
($q\bar{q}$) exchange mechanisms.  

The relevant diagrams at leading twist, which may 
involve both resonant and non-resonant channels, 
are shown in Fig.~\ref{fig:feynman}.
The Primakoff process $\gamma^\star \gamma^\star \to  \pp $
is not shown, because it is expected to contribute 
negligibly to the production of pions pairs with helicity 
zero or one~\cite{diehl_private}, and the analysis reported 
here is insensitive to helicity two.
Previous work~\cite{hermes_rho} has shown that 
resonant $\pp$ production via longitudinal 
$\rho^0$ decay in the kinematical region covered 
by the HERMES experiment occurs primarily through 
\mbox{two-quark} exchange  with the target.  
In the present more general case, the 
$q\bar{q}$ exchange mechanism gives rise to 
pion pairs with the values of the strong 
isospin $I$, total angular momentum $J$, 
and \mbox{$C$-parity} of either a \mbox{$\rho$-meson} 
\mbox{($I=1$, $J=1,3...,C=-1$)}, or an 
\mbox{$f$-meson} \mbox{($I=0$, $J=0,2...,C=+1$)}.
The $q\bar{q}$ exchange with $C=+1$ ($C=-1$) 
is described by flavour singlet (non-singlet) parton 
combinations~\cite{diehl}, and due to 
\mbox{$C$-parity} conservation the $\pi^+\pi^-$ 
pairs so formed have $C=-1$ ($C=+1$). 
The competing \mbox{two-gluon} channel gives rise 
to pion pairs with the quantum numbers of the 
$\rho$-meson family only. 
Pion pairs are formed from either quarks 
(\mbox{Fig.~\ref{fig:feynman}-a,b,c}) or 
gluons (\mbox{Fig.~\ref{fig:feynman}-d}) 
produced in the perturbative 
hard part of the reaction. 
Since the cross section for isovector $\pp$ production 
is much larger than for the isoscalar case, it is 
difficult to obtain experimental data on the isoscalar 
channel.  
One possible solution would be to study exclusive 
$\pi^0\pi^0$ production, but this requires a large 
experimental acceptance.  
With charged pions, the interference between the two 
isospin channels can also provide information on the 
weaker isoscalar channel at the amplitude level. 

For the purpose of studying the interference between 
$\pp$ production in $P$-wave \mbox{($I=1$)} 
and \mbox{$S,D$-wave} states \mbox{($I=0$)}, 
the Legendre moments 
$\langle P_1(\cos\theta) \rangle$ and  
$\langle P_3(\cos\theta) \rangle$ 
are particularly useful because they are sensitive 
only to such interference. 
The Legendre moment of order $n$ is given by 
\be
  \langle P_n(\cos\theta)\rangle^{\pp} =
  \frac{\int_{-1}^1 d \cos\theta \, P_n(\cos\theta)\,
  \frac{d\sigma^{\pp}}{d \cos\theta}}
  {\int_{-1}^1 d \cos\theta\,\frac{d 
  \sigma^{\pp}}{d \cos\theta}}\ , 
  \label{eq:int_density}
\ee
where $\theta$ is the polar angle of the $\pi^+$ meson 
with respect to the direction of the $\pp$ pair in the 
\mbox{center-of-momentum} frame of the virtual photon    
and target nucleon.
The moments $\la P_1 \ra$ and $\la P_3 \ra$ 
have been evaluated as a function of the
pion pair invariant mass $m_{\pi\pi}$, and
the Bjorken variable
\mbox{$x=\frac{Q^2}{2\nu M_P}$}, where $-Q^2$ is
the squared \mbox{four-momentum} of the initial
virtual photon, $M_P$ is the proton mass
and $\nu$ is the virtual photon energy
in the target rest frame.
Experimentally, $\la P_n \ra$ is the average 
of $P_n(\cos\theta_i)$ for all events $i$ 
grouped in bins of $m_{\pi\pi}$ or $x$.
%

In general, 
\be
\frac{d\sigma^{\pi^+\pi^-}}{d\cos\theta}
\propto\sum_{JJ^\prime \lambda\lambda^\prime}
\rho_{\lambda\lambda^\prime}^{JJ^\prime}
Y_{J\lambda}(\theta,\phi)
Y^\star_{J^\prime\lambda^\prime}(\theta,\phi) 
\ee
in which $\rho$ is the spin density matrix of the pion 
pair, whose diagonal entries $\rho_{\lambda\lambda}^{JJ}$ 
give the probability of producing it with angular momentum 
$J$ and longitudinal projection $\lambda$, and whose 
\mbox{off-diagonal} terms describe the corresponding 
interference terms.
If parity is conserved 
$\rho_{\lambda\lambda^\prime}^{JJ^\prime}$ 
is real and 
\mbox{$\rho_{\lambda\lambda^\prime}^{JJ^\prime}
=(-1)^{\lambda-\lambda^\prime}
\rho_{-\lambda-\lambda^\prime}^{JJ^\prime}$}~\cite{sekulin}. 
The contributions for \mbox{$J>2$}
are expected to be negligible in the
\mbox{$m_{\pi\pi}$-range} covered by HERMES.
The Legendre moments then are 
\begin{subequations}
\be
\la P_1 \ra & = & \frac{1}{\sqrt{15}} 
\left[ 4\sqrt{3}\rho^{21}_{11} + 4 \rho^{21}_{00} 
+2\sqrt{5}\rho^{10}_{00} \right] \ ,
\label{eq_p1}
\ee
\be
\la P_3 \ra & = & \frac{1}{7\sqrt{5}}
\left[ -12\rho^{21}_{11} + 6\sqrt{3} \rho^{21}_{00}
\right] \ .
\label{eq_p3}
\ee
\end{subequations}
In particular, $\la P_1 \ra$ 
is sensitive to \mbox{$P$-wave} interference with 
\mbox{$S$ and $D$-waves}, whereas $\la P_3 \ra $
is sensitive to only \mbox{$P$-wave} 
interference with a \mbox{$D$-wave}. 

The relevant factorization theorem~\cite{collins} has 
been proved only for longitudinal virtual photons 
$\gamma^*_L$ in leading twist.  
Contributions from transverse photons $\gamma^*_T$ 
and other \mbox{higher-twist} effects are suppressed by 
powers of $1/Q$. 
Therefore, the longitudinal terms $\rho^{21}_{00}$ 
and $\rho^{10}_{00}$ in \mbox{Eqs.~\ref{eq_p1} and~\ref{eq_p3}}
are expected to be dominant in the $m_{\pi\pi}$ 
region far from the $f_2$ meson, where the \mbox{higher-twist} 
term $\rho^{21}_{11}$ can be neglected.
On the other hand, in the region of the $f_2$ resonance 
the possible $\rho^{21}_{11}$ contribution can be 
eliminated by taking a combination of $\la P_1 \ra$ 
and $\la P_3 \ra$ that projects out the longitudinal 
terms:
\be
  \la P_1 + \frac{7}{3} P_3 \ra  & = &
  {2\sqrt{5\over3}}\rho_{00}^{21}
  + {2\over \sqrt{3}}\rho_{00}^{10} \ .
\ee
Assuming s-channel helicity conservation, such that 
the \mbox{0-helicity} photon $\gamma_L^*$ produces a 
$\pi^+\pi^-$ pair with \mbox{0-helicity}, only $\rho_{00}$ 
states are populated by $\gamma_L^*$.  
In this case, the combination
\mbox{$\la P_1 + \frac{7}{3} P_3 \ra$} would be 
sensitive to longitudinal photons only.
In the $f_2$ region, far from the $\rho^0$ and $f_0$
resonances, the term $\rho_{00}^{10}$ is expected to 
vary very slowly with $m_{\pi\pi}$, making no 
contribution to any structure appearing in this 
combination. 

In the $m_{\pi\pi}$ region of the $f_2$ meson,  
another combination eliminates the contribution 
of longitudinal tensor pairs:  
\be
  \la P_1 - \frac{14}{9} P_3 \ra & = &
  {4\sqrt{5}\over 3} \rho_{11}^{21}
  + {2\over \sqrt{3}}\rho_{00}^{10} \ .
\ee
Hence, the transverse \mbox{higher-twist} $\rho^{21}_{11}$ 
and longitudinal \mbox{leading-twist} $\rho^{21}_{00}$ 
contributions to the Legendre moments in the $f_2$ domain  
can be disentangled by comparing the combinations given above. 
%

%
\begin{figure*}[t!]
 \begin{center}
  \includegraphics[height=6.1cm,width=7.1cm]{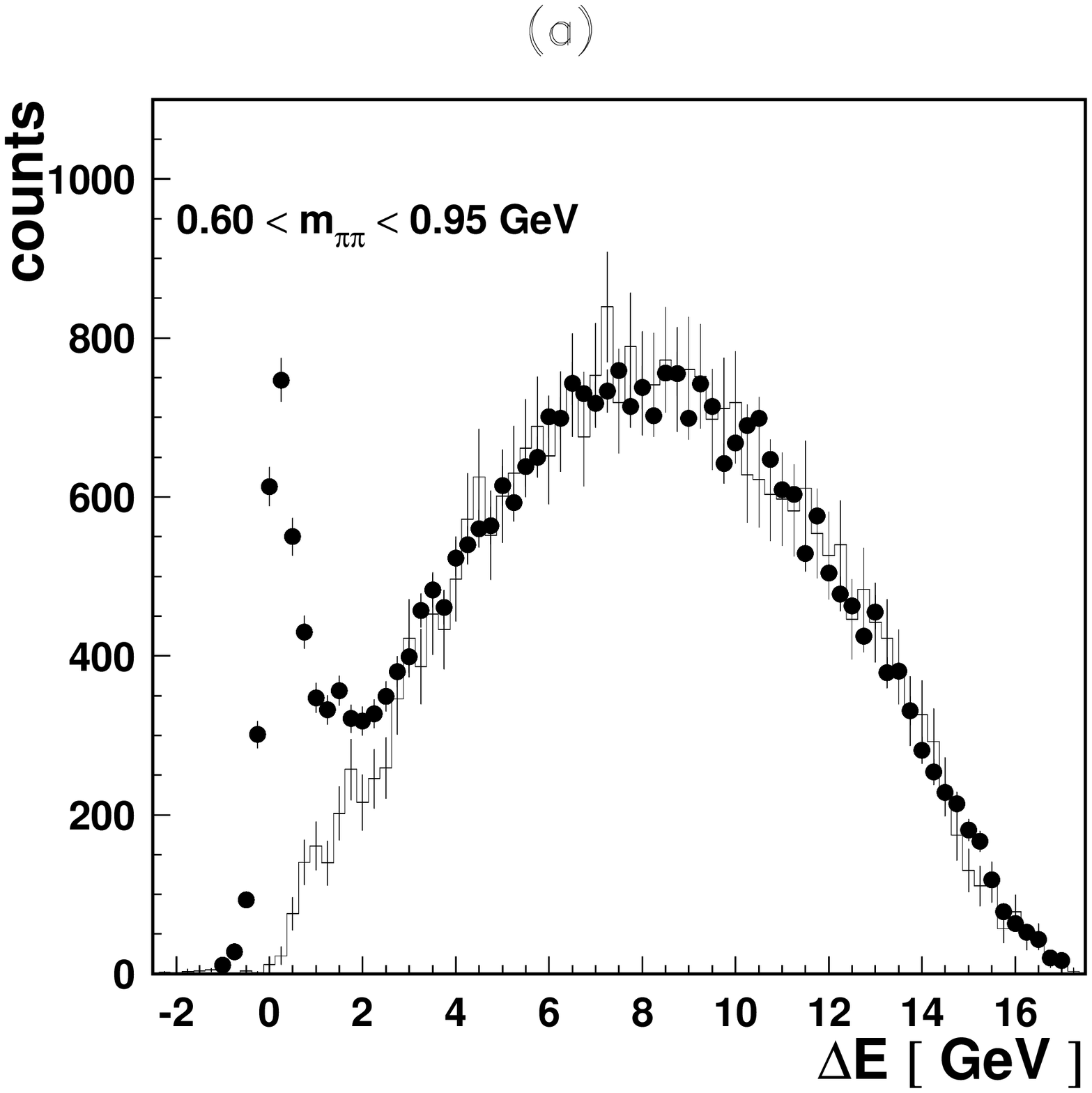} 
  \hspace{1.5cm}
  \includegraphics[height=6.1cm,width=7.1cm]{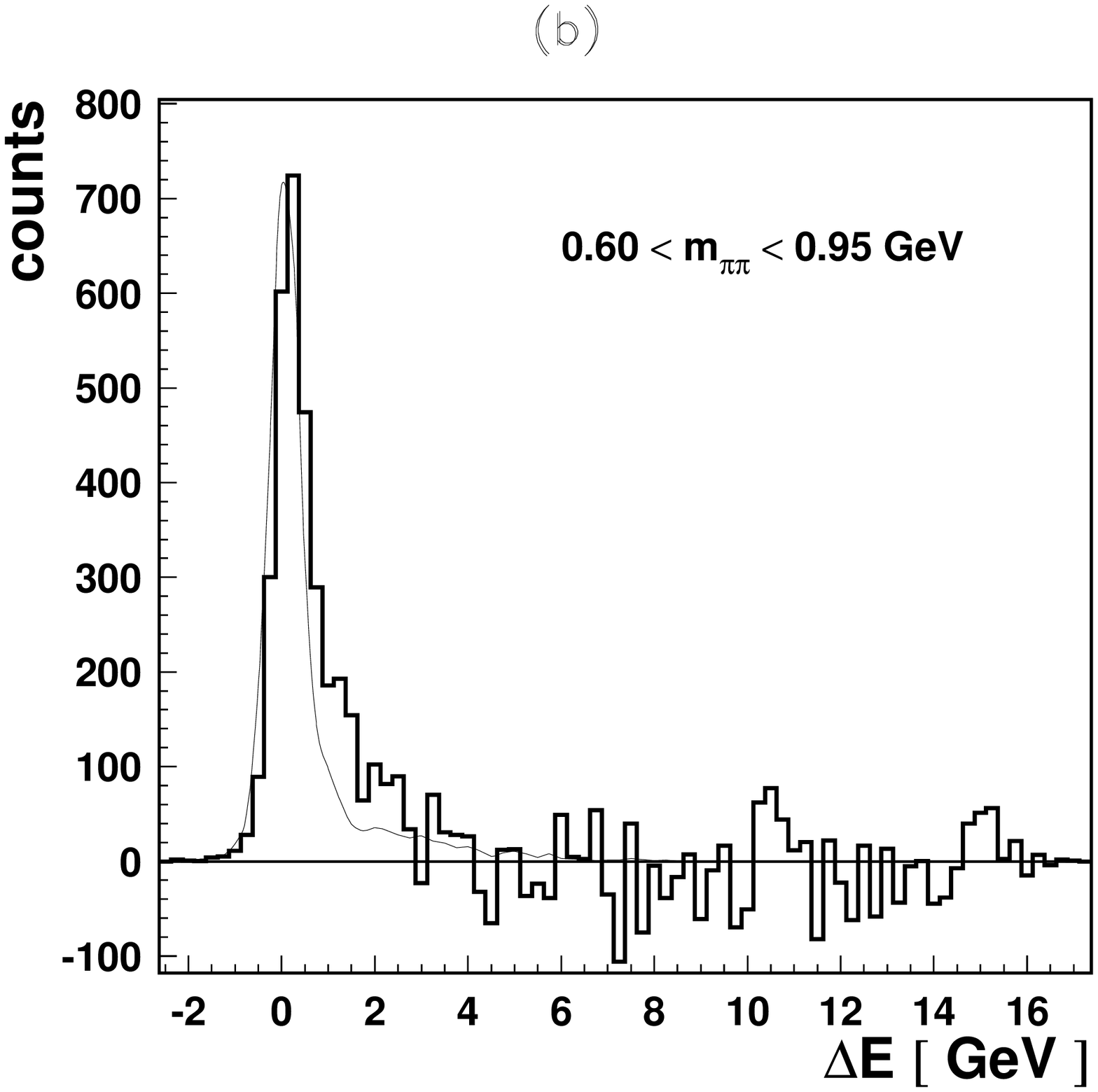}
  {\caption{Panel (a): Distribution of $\pi^+\pi^-$
            events versus $\Delta E$ for hydrogen 
            with \mbox{$0.60 < m_{\pi\pi} < 0.95$ GeV}. 
            The data are represented by the solid circles, 
            while the simulated (SIDIS) background is 
            represented by the histogram. 
            The Monte Carlo results are normalized to the data 
            using the region of the spectrum above 
            $\Delta E>2$ GeV. 
            Panel (b): 
            Yield of the exclusive events as obtained 
            by subtracting the normalized Monte Carlo events 
            from the data. 
            The result (thin line) of an arbitrarily normalized 
            Monte Carlo simulation using the diffractive 
            $\rho^0$ DIPSI generator is superimposed on the 
            exclusive distribution.
            }
  \label{fig:procedure}}
 \end{center}
\end{figure*}
The data were collected with the HERMES
spectrometer~\cite{spectrometer}
during the running period \mbox{1996-2000}.
The \mbox{27.6 GeV} HERA positron beam at DESY was
scattered off hydrogen and deuterium targets.
Events were selected with exactly one positron track
and two oppositely charged hadron tracks with
momentum \mbox{$>1$ GeV},
requiring that no additional neutral clusters
occur in the calorimeter.
Positrons were distinguished from hadrons with
an average efficiency of $98\%$, and a hadron
contamination below $1\%$, over the whole
kinematic range.
In order to ensure a hard scattering  process,
the constraints \mbox{$Q^2>1$ GeV$^2$} and
\mbox{$W>2$ GeV} were imposed, where $W$ is 
the invariant mass of the virtual photon-nucleon 
system.

When studying the $m_{\pi\pi}$-dependence 
of the Legendre moments, the requirement    
\mbox{$x>0.1$} was imposed to suppress 
the contribution from gluon-exchange relative 
to that from $q\bar{q}$ exchange~\cite{polyakov01}. 
However, when analyzing
the $x$-dependence of the Legendre moments,
the whole \mbox{$x$-range} accessible to 
\mbox{HERMES} was used.

Since the recoiling target nucleon is not detected in 
the present HERMES apparatus, exclusive events 
were selected by restricting the quantity 
$\Delta E = \frac{M^2_X - M^2_{targ}}{2M_{targ}}$, 
in which $M_X$ is the missing mass, and $M_{targ}$  
is the nucleon target mass. 
A $\Delta E$ distribution peaked at zero 
is a clear signature of exclusive production, 
while larger $\Delta E$ values indicate 
non-exclusive events.
For scattering off nuclei, one can have 
either incoherent scattering from individual 
nucleons inside the target \mbox{($M_{targ}\approx M_N$)} 
or coherent scattering from the entire nucleus $A$ 
\mbox{($M_{targ}\approx M_A$).} 
For scattering off deuterium, incoherent scattering
is found to dominate for HERMES kinematics~\cite{rho_coherent}; 
therefore $M_{targ}$ was chosen to be the proton
mass throughout the entire analysis.  
All detected hadrons have been treated as pions.
%

In the $\Delta E$ spectrum, the resolution 
due to instrumental effects ranges between 
0.260 and 0.380 GeV, depending 
on the data production year.  
Thus, even at low $\Delta E$ the sample is 
contaminated by non-exclusive processes. 
This background yield was assumed to be 
semi-inclusive deep inelastic scattering 
(SIDIS) events and was evaluated by first 
calculating the $\Delta E$ distribution of 
SIDIS events with a Lepto \mbox{Monte Carlo} 
simulation~\cite{lepto,lund},  
and then normalizing it to the data in the range 
\mbox{$\Delta E > 2$} GeV.
The effect of varying this normalization 
region was treated as a systematic uncertainty contribution. 
Fig.~\ref{fig:procedure} shows the normalized 
\mbox{Monte Carlo} distribution in $\Delta E$ 
compared to the data, and their difference. 
The simulated background shape is in agreement 
with the data at large $\Delta E$, while at 
small $\Delta E$ the data show a surplus 
due to the presence of the 
exclusive process not included in that  
Monte Carlo simulation. 
Comparison of the exclusive peak in 
the data with the result of a \mbox{Monte Carlo} 
simulation using the diffractive $\rho^0$ DIPSI 
generator~\cite{dipsi} reveals an excess at 
$\Delta E \approx 1.5$ GeV. 
This excess can be explained by the combined contributions of 
$\rho^0$ production via single and double-dissociation
of the proton as described in Ref.~\cite{hermes_rho2}, 
and of radiative corrections~\cite{elke}, which all three 
are not simulated by the DIPSI Monte Carlo.
\begin{figure*}[t!]
 \begin{center}
  \includegraphics[height=6.1cm,width=7.1cm]{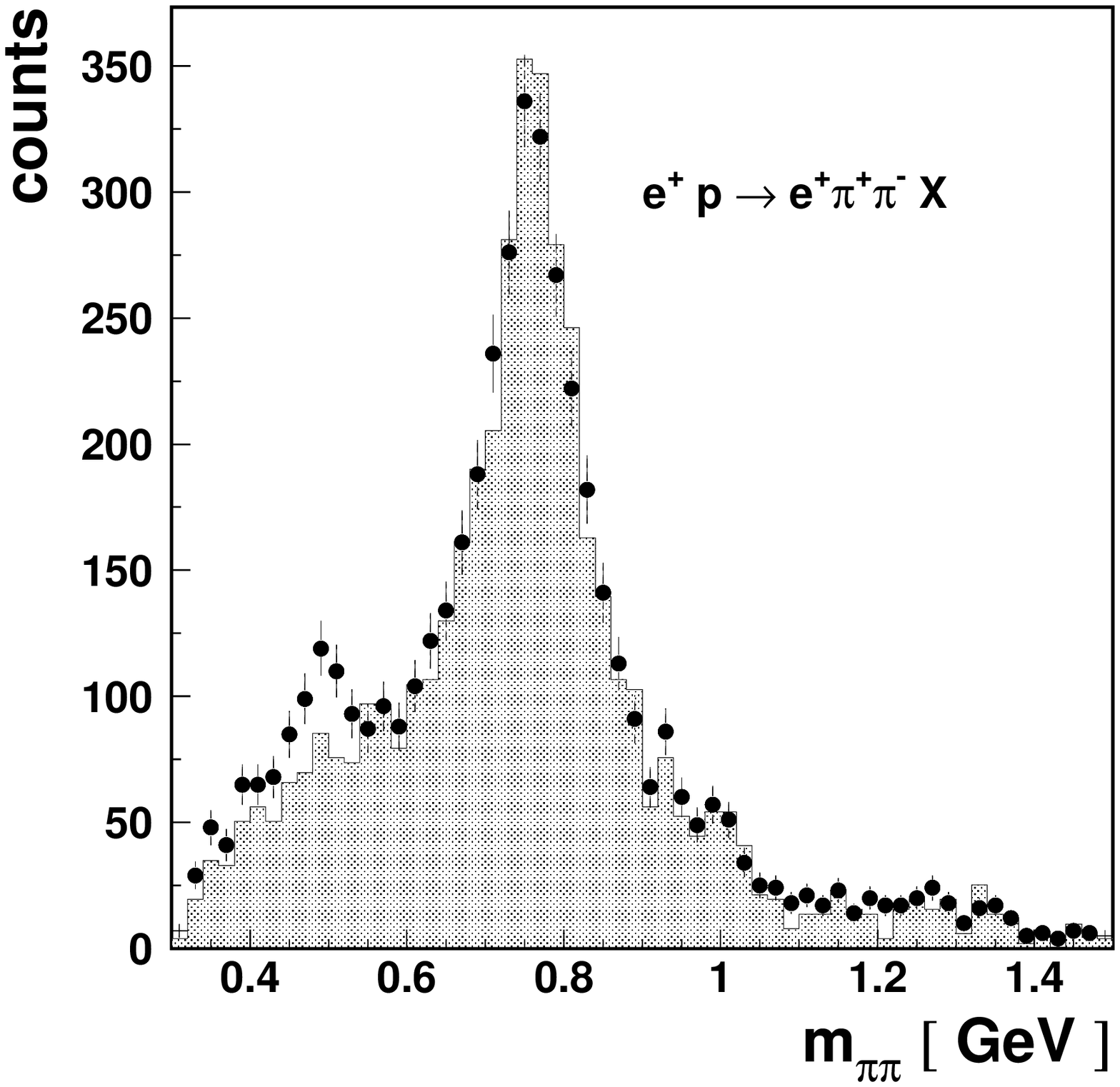}
  \hspace{1.5cm}
  \includegraphics[height=6.1cm,width=7.1cm]{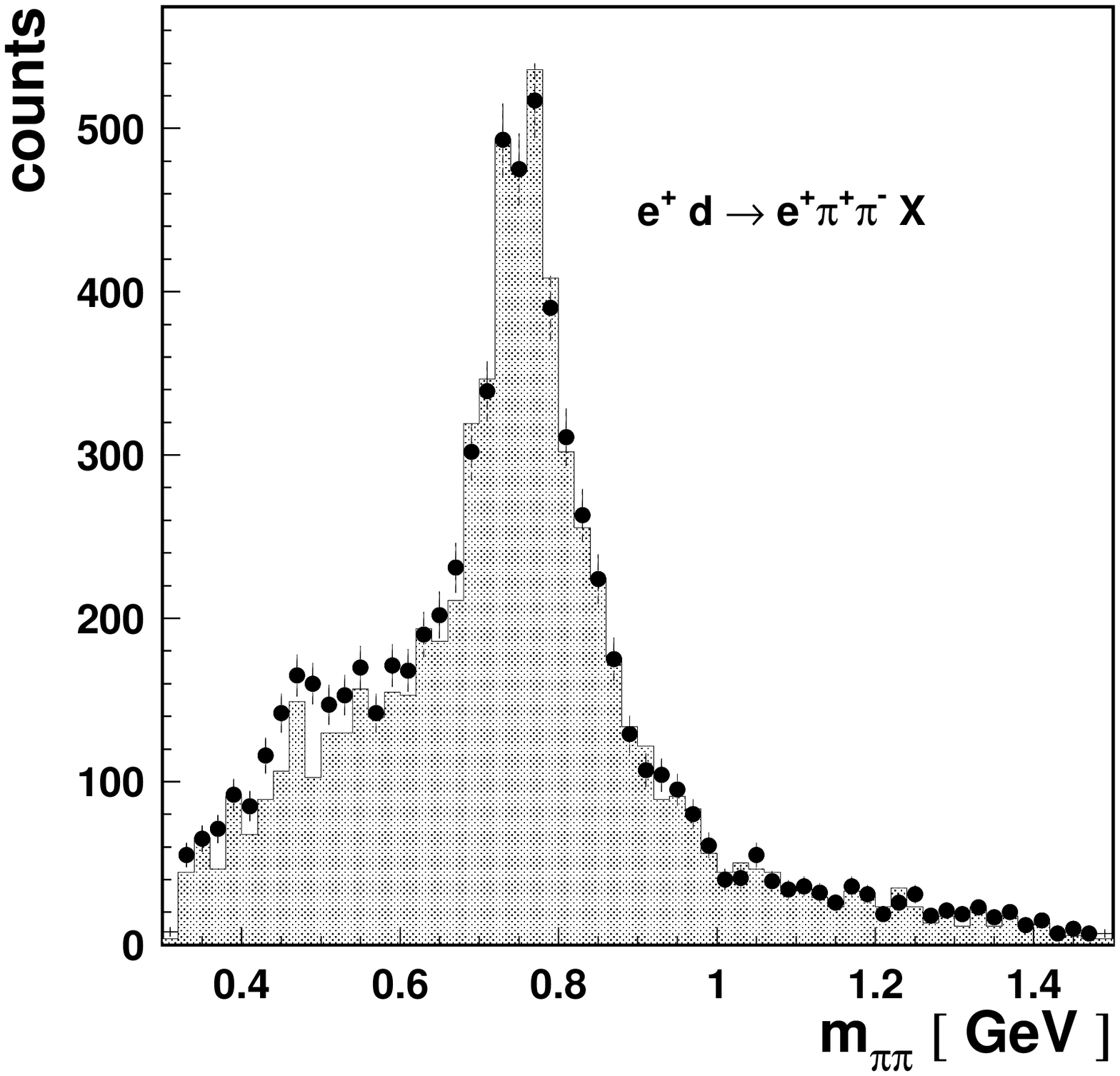}
  {\caption{Invariant mass spectrum for hydrogen (left)  
            and deuterium (right) for $\Delta E<0.625$ GeV
            (solid points) and \mbox{$\Delta E<0.125$ GeV} 
            (shaded area). 
            For both spectra, the requirement $x>0.1$
            has been applied.
            For both targets, the \mbox{$m_{\pi\pi}$-spectrum} 
            for $\Delta E<0.125$ GeV 
            is normalized and superimposed (shaded area) 
            to show the suppression of 
            the $\omega \rightarrow \pi^+\pi^-\pi^0$ 
            contamination described in the text.
            }
  \label{fig:mass_spectrum}}
 \end{center}
\end{figure*}

In order to evaluate the background contribution 
to the exclusive signal, the experimental and the 
normalized \mbox{Monte Carlo} yields were separately 
integrated up to a limiting $\Delta E$ value 
$\Delta E_{cut}$, resulting in $N_{tot}$ and $N_{MC}$ 
respectively.  
The value of $\Delta E_{cut}$ was optimized 
by requiring the ratio of the exclusive signal 
\mbox{$N_{Sg}=N_{tot}-N_{MC}$} 
over the background ($N_{Sg}/N_{Bg}$) to be large, and 
the relative statistical uncertainty $\Delta N_{Sg}/N_{Sg}$ 
to be small.
The optimized $\Delta E_{cut}$ value  
for both targets is $0.625$ GeV. 
Below the chosen $\Delta E_{cut}$ value, the 
SIDIS contamination is found to range between 
$2\%$ and $65\%$ of the total events, depending 
on $m_{\pi\pi}$ and $x$. 
In particular, this contamination is small 
at $m_{\pi\pi}$ values around $m_{\rho^0}$, 
and increases at smaller and larger 
invariant mass values. 

The SIDIS model does not account for contamination 
from other processes. 
In order to suppress the 
$\omega \rightarrow \pi^+\pi^-\pi^0$ decay  
at low $m_{\pi\pi}$, as explained below, 
a more severe $\Delta E_{cut}$ was applied  
than the value optimized for the SIDIS background. 
The final $\Delta E_{cut}$ values used in this 
analysis for both targets are $0.125$ GeV for 
\mbox{$m_{\pi\pi}\le 0.60$ GeV}, and $0.625$ GeV for 
\mbox{$0.60<m_{\pi\pi}\le 1.40$ GeV}. 

The limited $\Delta E$ resolution does not 
allow for the complete suppression of single 
and \mbox{double-dissociation} processes.
An example is the process in which the nucleon is left in
a $\Delta$ resonance state that decays 
with an unobserved pion. 
The contamination from single and 
\mbox{double-dissociation} 
was estimated by shifting the value of 
$\Delta E_{cut}$ by \mbox{0.5 GeV}, 
from a low value of \mbox{0.125 GeV}
where this contamination is negligible, to a
relatively large value, \mbox{0.625 GeV},
where this background is possibly substantial. 
This effect was included in the systematic
uncertainty.

The contamination from target excitations such as
$e^+ p \rightarrow e^+ \pi \Delta \rightarrow e^+ p \pi^+ \pi^-$,
which have been found to contaminate the process
$e^+ p \rightarrow e^+ p \pi^+ \pi^-$ at lower energy 
and $W$ values~\cite{Clas}, in the HERMES kinematics
were found to be negligible in a \mbox{Dalitz-plot} 
analysis~\cite{thesis}.

The contamination of exclusive $K^+K^-$ 
pairs from $\phi(1020)$ meson decay, which 
appears in the event yield at 
\mbox{$m_{\pi\pi} \approx 0.35$} GeV, is entirely 
eliminated by applying the additional 
cut \mbox{$m_{KK} > 1.06$ GeV}. Here $m_{KK}$ 
is the invariant mass of the two hadrons 
when they are treated as kaons.
\begin{figure*}[htb!]
\begin{center}
 
\hspace{0.90cm}
  \includegraphics[height=7.6cm,width=7.2cm]{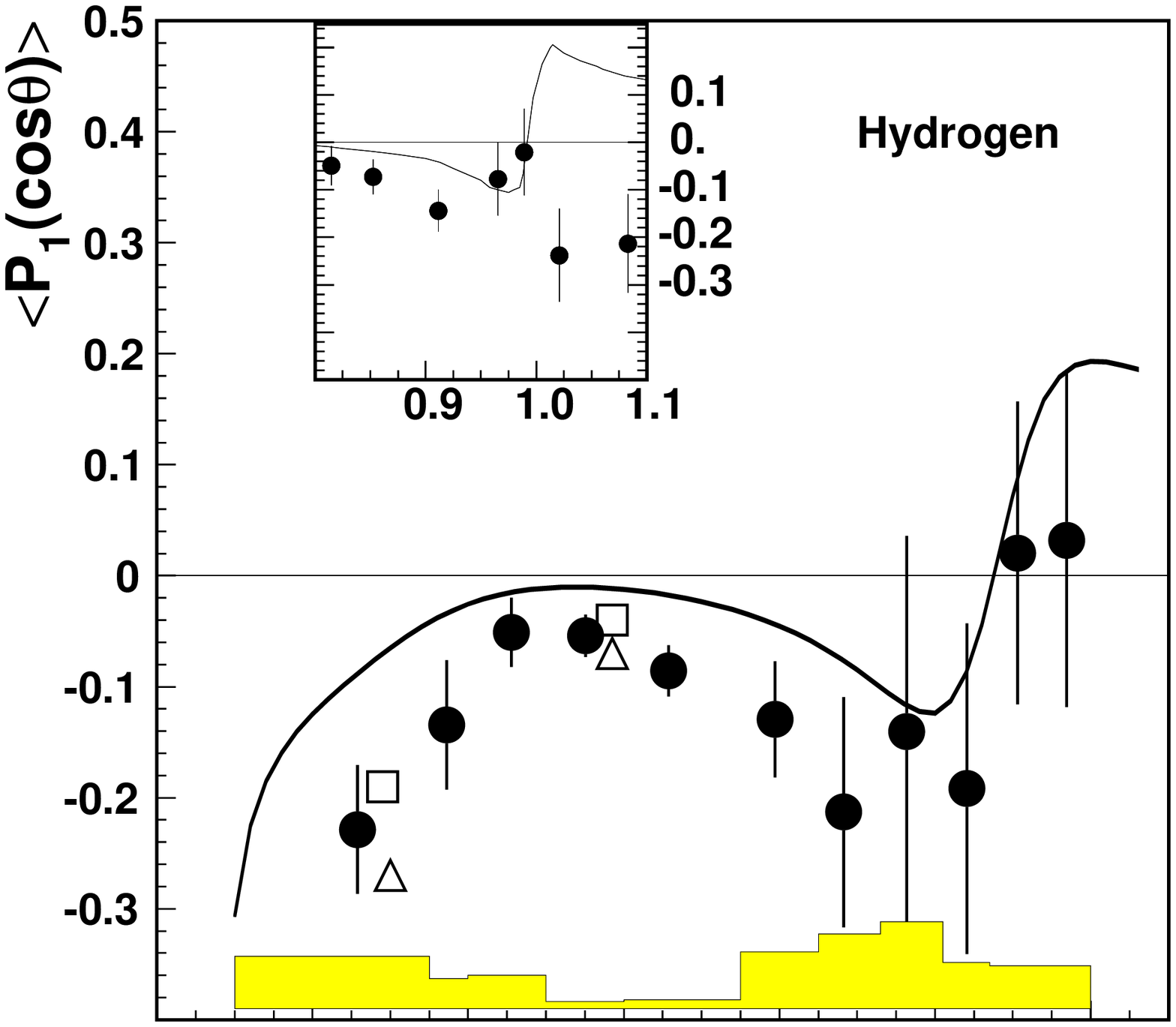}
\hspace{0.80cm}
  \includegraphics[height=7.6cm,width=7.2cm]{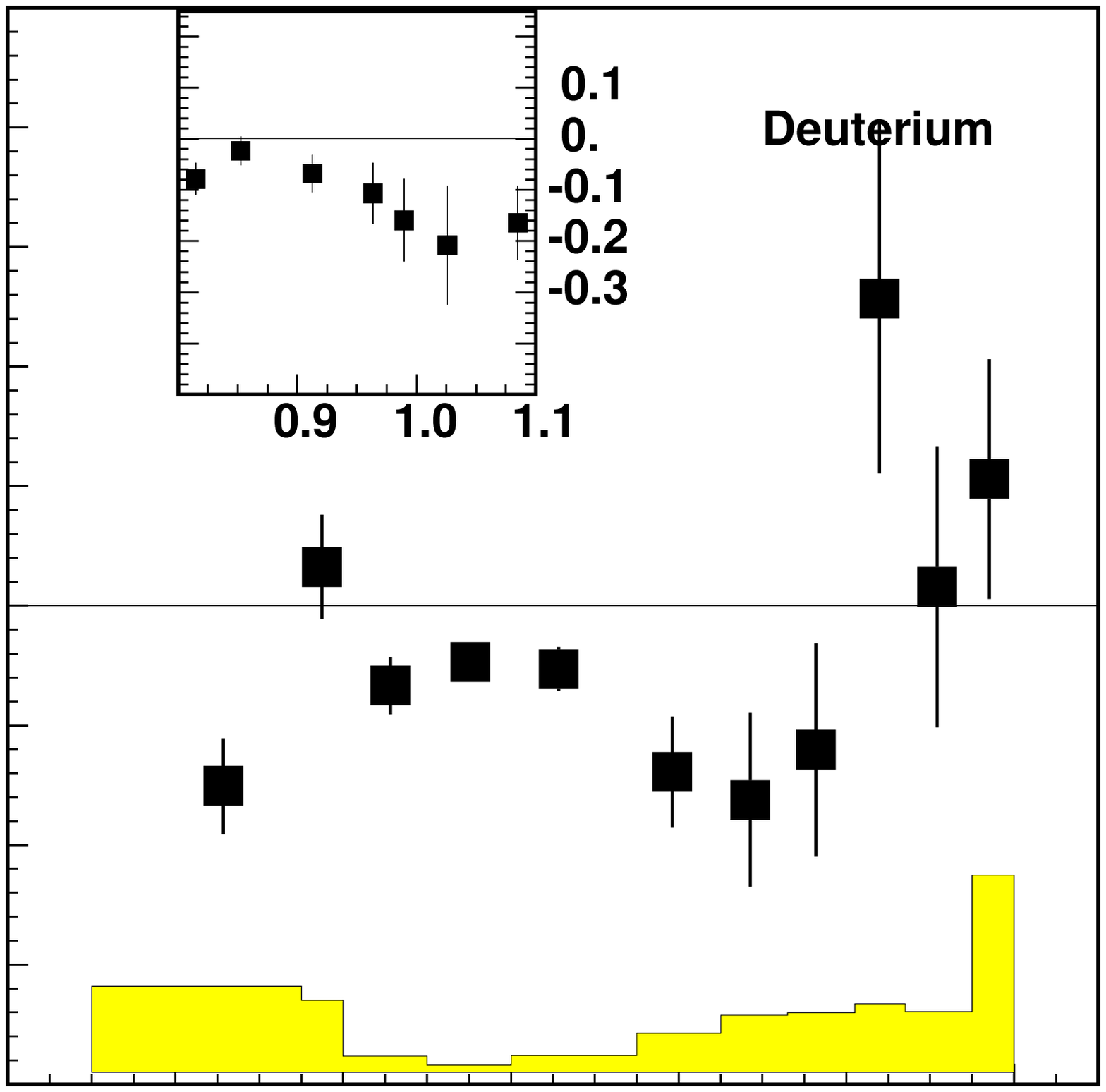} \\
 
  \vspace{-1.65cm}
\hspace{0.90cm}
  \includegraphics[height=7.6cm,width=7.2cm]{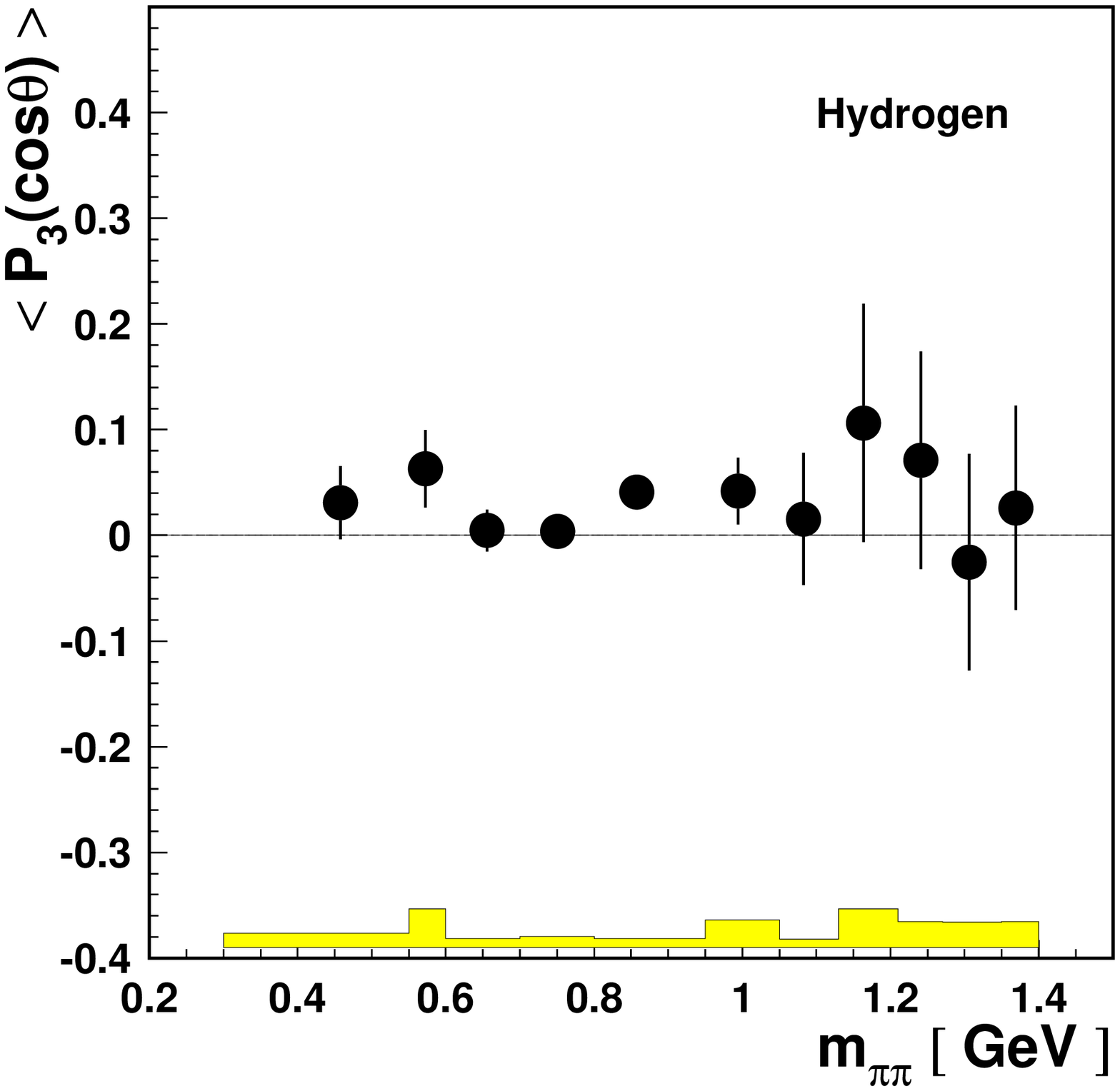}
\hspace{0.80cm}
  \includegraphics[height=7.6cm,width=7.2cm]{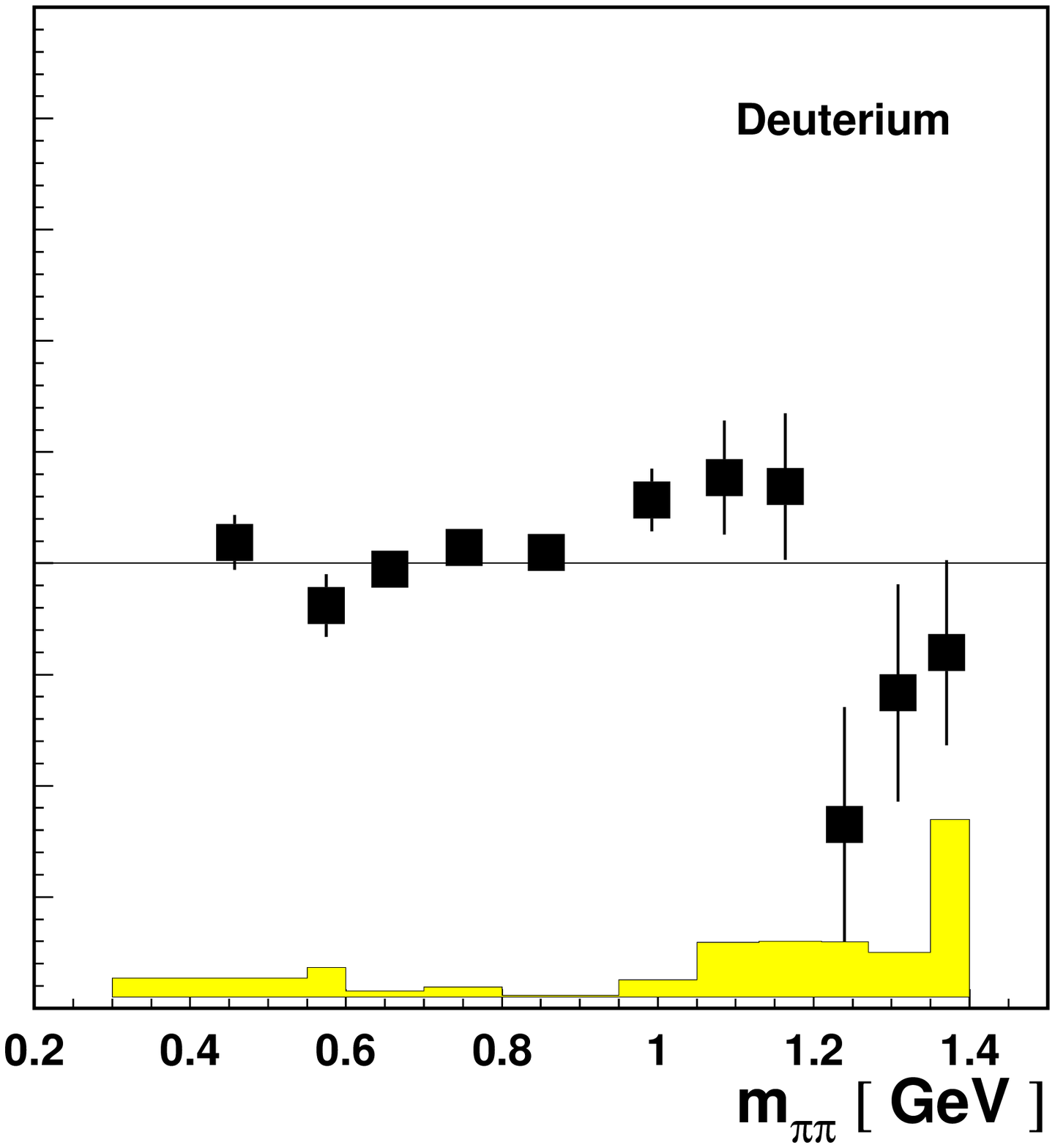}
  {\caption{\mbox{The $m_{\pi\pi}$-dependence} of the
            Legendre moments
            $\la P_1 \ra$ (upper panels) and
            $\la P_3 \ra$ (lower panels)
            for hydrogen (left panels) and 
            deuterium (right panels), 
            for $x>0.1$. 
            The region
            $0.8<m_{\pi\pi}<1.1$ GeV is
            presented with finer bins 
            to better investigate possible
            contributions from the narrow
            $f_0(980)$ resonance,
            as shown in the insert.
            In the upper panels, leading twist predictions for
            the hydrogen target including
            the two-gluon exchange mechanism contribution,
            LSPG~\cite{polyakov01,lehmann}
            (solid curve) at $x=0.16$ are shown.
            A calculation without the gluon exchange contribution
            is shown for limited $m_{\pi\pi}$ values,
            LPPSG~\cite{polyakov00}
            (open squares at $x=0.1$, open triangles at
            $x=0.2$).
            In these calculations, the contribution
            from $f_0$ meson decay was not considered.
            Instead, the inset panel for the hydrogen
            target shows the prediction from~\cite{pire},
            which includes the $f_0$ meson contribution. 
            All experimental data have $\la x \ra =0.16$, 
            \mbox{$ \la Q^2 \ra$ = 3.2 (3.3) GeV$^2$},
            and \mbox{$\la -t \ra = 0.43$ (0.29) GeV$^2$} 
            for hydrogen (deuterium). 
            The systematic
            uncertainty is represented by the error band.}
            \label{fig:mpp_results}}
\end{center}
\end{figure*}
Similarly, the contamination of 
$\phi\rightarrow K_SK_L$, with $K_S$ detected through 
its decay in $\pi^+\pi^-$, by using a \mbox{Monte Carlo} 
DIPSI simulation was found to be entirely 
absent within the chosen $\Delta E_{cut}$ values.
The channel $\omega \rightarrow \pi^+\pi^-$ 
and exclusive \mbox{non-resonant} 
$K^+K^-$~{$^($\footnote{
This contamination has been estimated by comparing 
results from the data and the Monte Carlo simulation 
of SIDIS events. }$^)$} production were estimated to 
contaminate the signal by less than  \mbox{$0.3\%$} 
and \mbox{$1.5\%$}, respectively, and were neglected.
The decays 
$\phi \rightarrow \pi^+\pi^- \pi^0$~{$^($\footnote{
Including the resonant channel 
$\phi \rightarrow \rho \pi \rightarrow \pi^+\pi^- \pi^0.$
}$^)$},
with the $\pi^0$ outside the acceptance,
gives a contamination of less than \mbox{$1\%$}.
A contamination of about $18\%$ from the decay 
$\omega \rightarrow \pi^+\pi^-\pi^0$, 
with only the charged tracks detected, yields 
a reconstructed $m_{\pi\pi}$ distribution
centered at $0.45$ GeV with a Gaussian width of
approximately $0.075$ GeV~\cite{tytgat}. 
This contribution to the yield was suppressed 
by imposing \mbox{$\Delta E < 0.125$ GeV} 
in the region \mbox{$m_{\pi\pi} \le 0.6$ GeV}.   
The effect of the remaining contamination 
was taken into account in the systematic 
uncertainty of the relevant bins.
All the above estimations of these additional 
background components are small compared to the 
background predicted by the SIDIS model. 

After applying all event selection requirements, 
\mbox{$4.8 \times 10^3$} (\mbox{$7.2 \times 10^3$})
$\pi^+\pi^-$ events remained for the
\mbox{$m_{\pi\pi}$-dependence} analysis
with $x>0.1$,
and $11.0 \times 10^3$ ($13.3 \times 10^3$)
events for the \mbox{$x$-dependence} analysis 
for hydrogen (deuterium).
The invariant mass spectra for hydrogen
and deuterium with \mbox{$\Delta E<0.625$ GeV}, 
$x>0.1$, and $m_{KK}>1.06$ GeV 
are shown in Fig.~\ref{fig:mass_spectrum}. 
%

In each of the analyzed bins, $\la P_n \ra_{data}$ was 
evaluated within the chosen exclusive $\Delta E$ region, 
with no background subtraction. 
The values of $\la P_n \ra_{SIDIS}$ for the background 
events were extracted from the data for 
\mbox{$\Delta E>2$ GeV}, where SIDIS events dominate. 
These values were found to be consistent when evaluated 
in three different $\Delta E$ bins: 
\mbox{$2< \Delta E< 4$ GeV}, \mbox{$4< \Delta E< 6$ GeV},
and \mbox{$ \Delta E > 6$ GeV}. 
The moments were corrected for SIDIS background using 
\be
 \la P_n \ra_{\mathrm{exclusive}} = 
 \frac{1+r}{r} \la P_n \ra_{\mathrm{data}} -
 \frac{1}{r} \la P_n \ra_{\mathrm{SIDIS}} \ ,
\ee
in which $r$ is the ratio of integrated exclusive data 
to background Monte Carlo events for $\Delta E < \Delta E_{cut}$
in the analyzed bin. 
%

A Monte Carlo generator based on the GPD framework 
for the hard $\pp$ exclusive process does not exist. 
Therefore the DIPSI generator was used to evaluate 
the effects of geometric acceptance and instrumental 
smearing on the Legendre moments, 
which were both found to be negligible~\cite{thesis}. 
This \mbox{Monte Carlo} simulation is in good 
agreement with the kinematic distributions of exclusive 
$\rho^0$ mesons observed at HERMES. 
%

The analyzed moments might be sensitive to radiative corrections 
that affect the $\cos\theta$ angular distribution.
For $\rho^0$ decay, which dominates in the cross section 
for exclusive $\pi^+\pi^-$ production, 
the angular distribution depends linearly 
only on the vector spin density matrix element $r_{00}^{04}$.  
In previous work~\cite{akushevich} the relative 
correction of $r_{00}^{04}$ for radiative corrections 
has been evaluated, and found to be less than 
$0.3\%$ at $\langle Q^2 \rangle \approx 3$ GeV$^2$ 
in the kinematics of the H1 and ZEUS experiments. 
At larger $x$, where the \mbox{HERMES} analysis is performed,
they are even smaller.
As a result of these considerations, radiative corrections 
effects have been neglected in this analysis.  

The $m_{\pi\pi}$-dependence of $\la P_1 \ra$ and
$\la P_3 \ra$ for exclusive $\pi^+\pi^-$ production off 
hydrogen and deuterium is presented in 
Fig.~\ref{fig:mpp_results}, for $x>0.1$.  
The average values of $Q^2$, $-t$, and $x$  
for both targets in this domain are reported 
in Tab.~\ref{tab_values}. 
For \mbox{$m_{\pi\pi} < 1$ GeV}, the moments are similar 
for the two targets. 
In each panel for $\langle P_1 \rangle$, the region 
\mbox{$0.8<m_{\pi\pi}<1.1$} GeV is shown as an insert 
with finer binning to better investigate possible 
contributions from the narrow $f_0(980)$ resonance.    

The values for $\la P_1 \ra$ differ significantly 
from zero, and depend strongly on $m_{\pi\pi}$.  
At small invariant mass, i.e. close to the threshold 
$2m_\pi$, this non-zero moment is interpreted as 
originating from the interference between  the lower 
tail of the isovector $\rho^0(770)$ (\mbox{$P$-wave}) 
with the \mbox{$S$-wave} \mbox{non-resonant} 
$\pi^+\pi^-$ amplitude.
At $m_{\pi\pi}$ values around $m_{\rho^0}$, the 
absolute value of this quantity shows a minimum, 
which is explained in terms of the overwhelming 
dominance of $\rho^0$ vector meson production 
in the denominator of the moment. 
The increase of the size of $\la P_1 \ra$ 
at larger invariant mass is due to 
the interference of the upper tail of the 
$\rho^0$ with the non-resonant $\pi^+\pi^-$ 
\mbox{$S$-wave} production. 
At $m_{\pi\pi}\approx 1$ GeV,
the observed oscillation in hydrogen $\la P_1 \ra$ 
suggests an interference between the
$\rho^0$ tail and the \mbox{$S$-wave} $\pi^+\pi^-$
production from the narrow $f_0$(980) resonance.
Moreover, in the  $f_2(1270)$ meson region, 
the data suggest a sign change 
caused by the interference between  
the $\rho^0$ upper tail and the $f_2$ 
(\mbox{$D$-wave}).

The Legendre moment $\la P_3 \ra$ is sensitive 
only to the interference of \mbox{$P$-wave} and 
\mbox{$D$-wave} states in $\pi^+\pi^-$ production. 
Consistent with the expectation that no resonance 
decay into $\pi^+\pi^-$ pairs in $D$-wave states
occurs for \mbox{$m_{\pi\pi}\leq 1$ GeV}, no 
interference is observed in this invariant mass region.
The $\la P_3 \ra$ moment for deuterium 
increases in magnitude in the $f_2(1270)$ meson region. 
A sign change is also prominently visible,
reflecting the interference of the \mbox{$P$-wave}
and \mbox{$D$-wave} resonant  $\pi^+\pi^-$ channels.
On the other hand, no such signature is evident 
in the hydrogen data. 
\begin{figure}[h!]
\begin{center}
 
\hspace{0.90cm}
  \includegraphics[height=5.3cm,width=3.55cm]{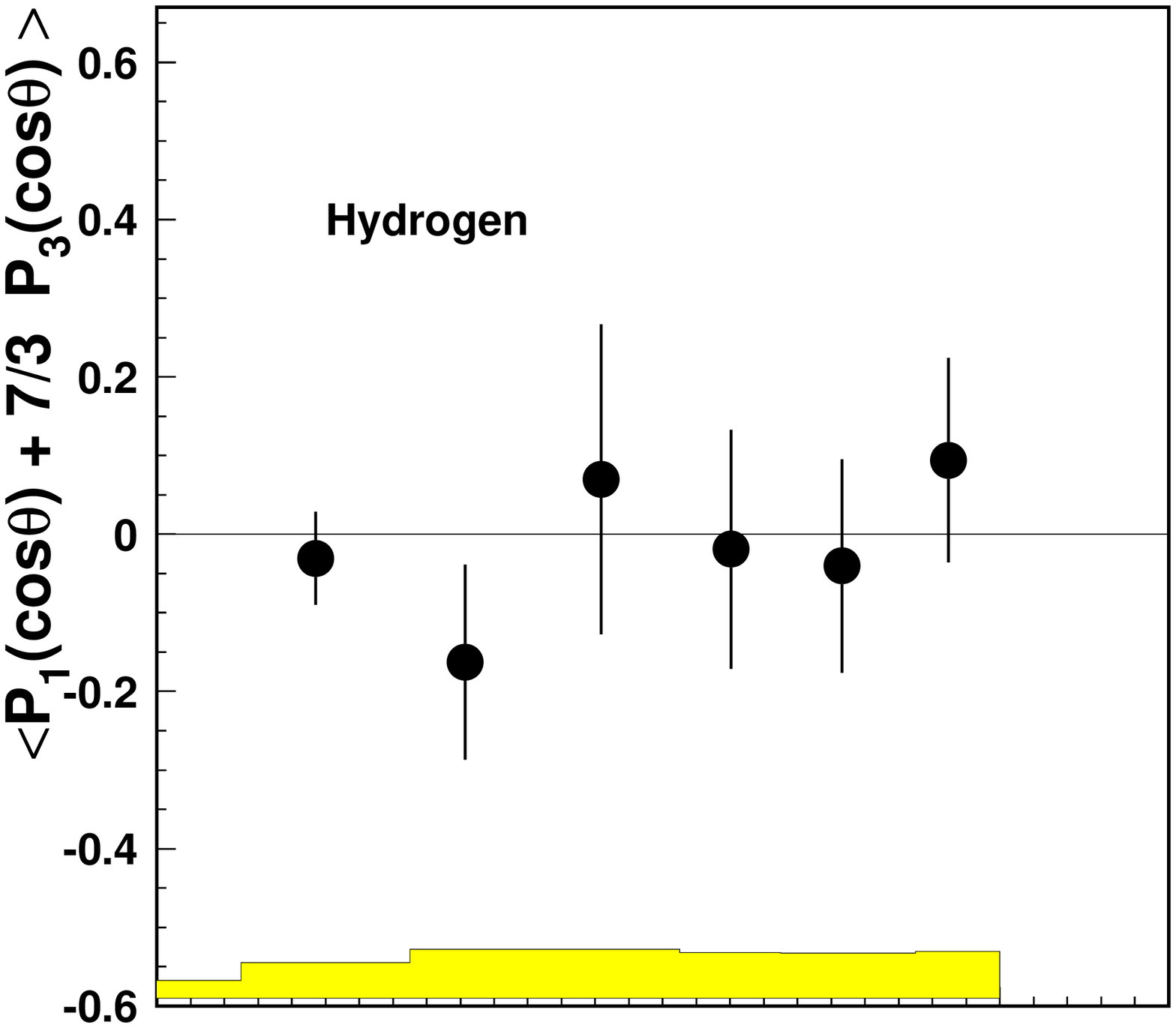}
\hspace{0.31cm}
  \includegraphics[height=5.3cm,width=3.55cm]{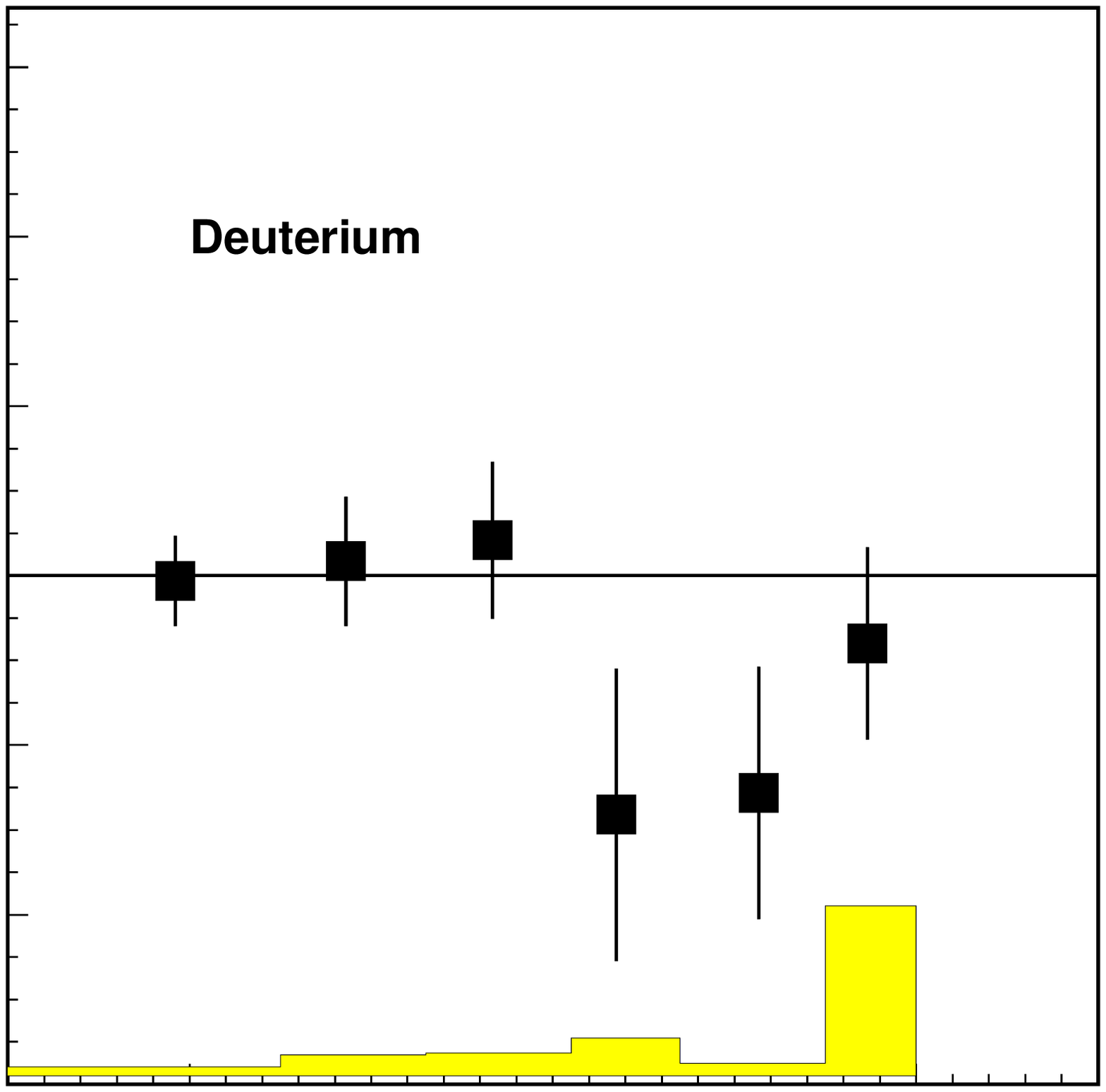} \\
 
  \vspace{-1.15cm}
\hspace{0.90cm}
  \includegraphics[height=5.3cm,width=3.55cm]{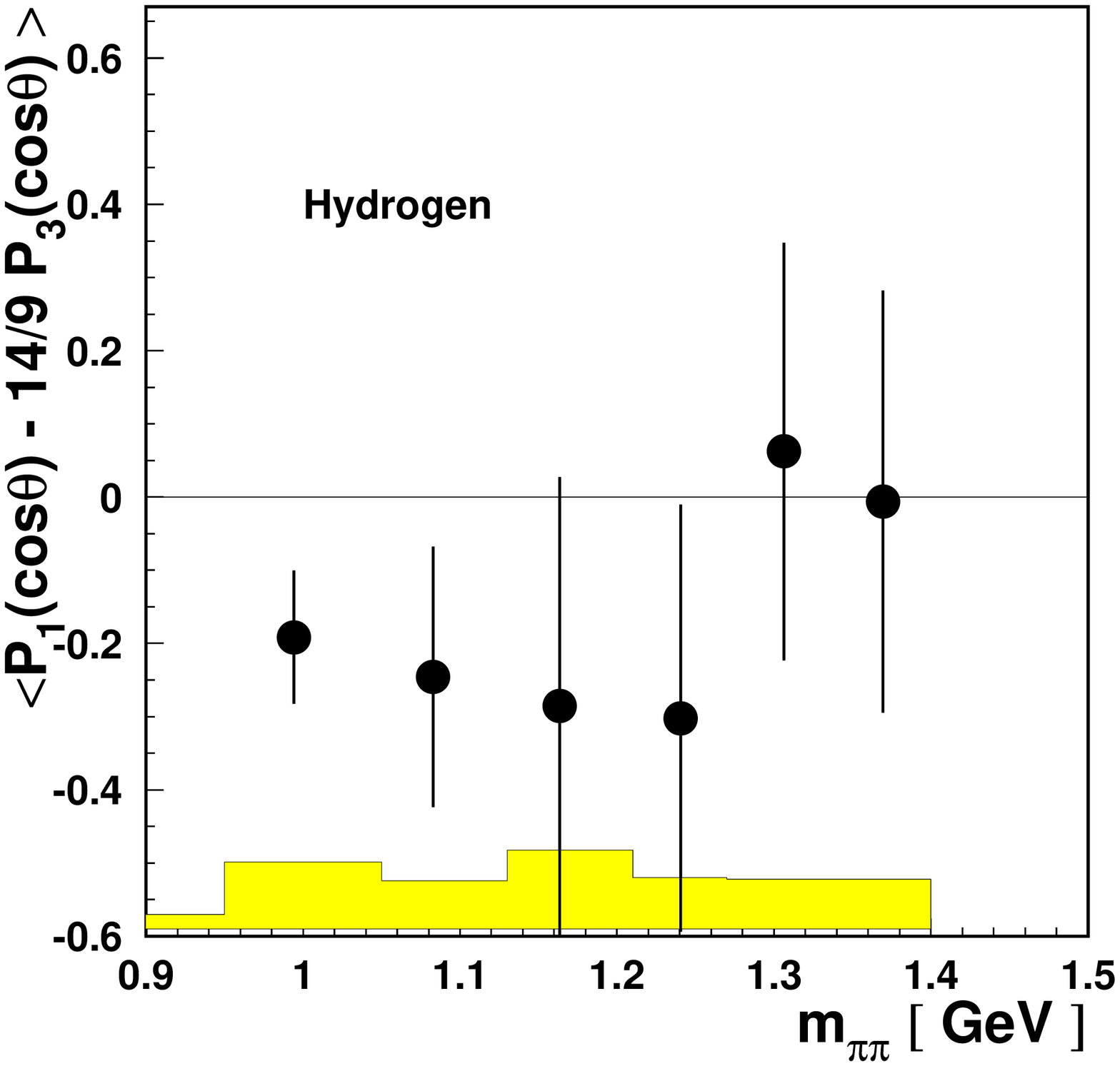}
\hspace{0.31cm}
  \includegraphics[height=5.3cm,width=3.55cm]{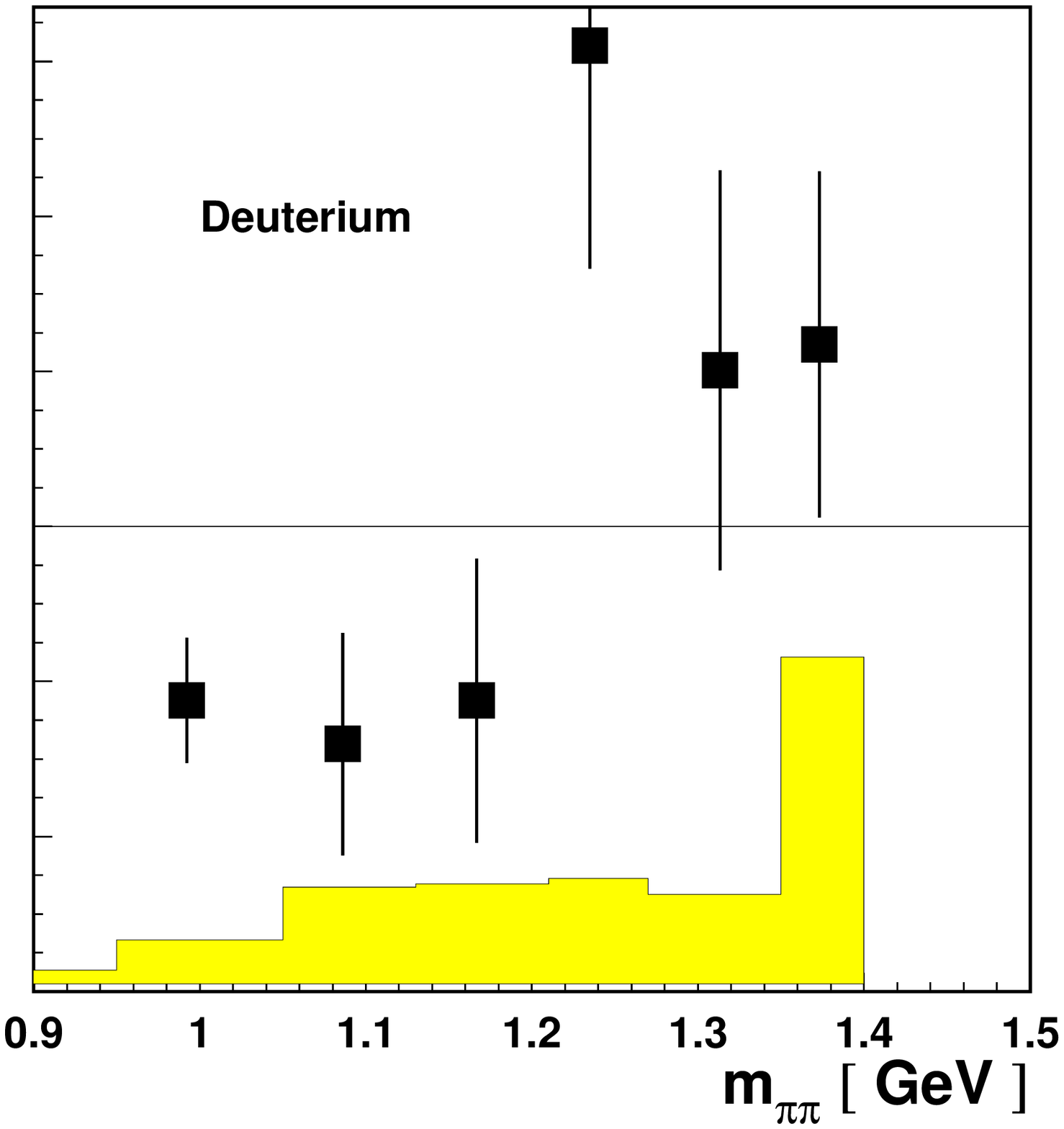}
  {\caption{The \mbox{$m_{\pi\pi}$-dependence} of 
            $\la P_1 + 7/3 \cdot P_3 \ra$ (upper panels)
            and $\la P_1 - 14/9 \cdot P_3 \ra$ (lower panels)
            for hydrogen (left panels) and deuterium
            (right panels).
            The data have $\la x \ra =0.16$,
            \mbox{$ \la Q^2 \ra$ = 3.2 (3.3) GeV$^2$},
            and \mbox{$\la -t \ra = 0.43$ (0.29) GeV$^2$}
            for hydrogen (deuterium).
            The systematic
            uncertainty is represented by the error band.}
            \label{fig:mpp_results_combinations}}
\end{center}
\end{figure}

In Fig.~\ref{fig:mpp_results} the
{\mbox{$m_{\pi\pi}$-dependence} of $\la P_{1} \ra$
for hydrogen is compared with theoretical
calculations based on the GPD framework,
with~\cite{polyakov01,lehmann} (solid curve)
and without~\cite{polyakov00} (open points)
the inclusion of the \mbox{two-gluon} exchange mechanism.
A possible contribution from the $f_0$ meson was
not considered in the calculations.
The calculations include only the longitudinal component
$\sigma_L$ of the $\pp$ cross section, while in this
analysis no separation between the $\sigma_L$ and
$\sigma_T$ contributions could be made.
The $\sigma_T$ contribution to the total cross section
for $\rho^0$ production is estimated to be
approximately $60\%$~\cite{hermes_rho2}.
The reasonable agreement of the leading twist
predictions for the $m_{\pi\pi}$-dependence of
the $\la P_{1} \ra$
data may tentatively be understood as arising from the
cancellation of higher twist effects in this
moment~\cite{garcon}.

To date, the $f_0$ contribution is taken into account
only by Ref.~\cite{pire}, where the discussion is restricted
to diffractive physics at \mbox{center-of-mass} energies
larger than \mbox{$100$ GeV}.
Nevertheless, to demonstrate the possible effect of this 
resonance, the comparison with those predictions for
$\la P_{1} \ra$ on hydrogen is shown in the panel insert
of Fig.~\ref{fig:mpp_results}.
\begin{figure*}[t!]
 \begin{center}

\hspace{0.90cm}
  \includegraphics[height=7.8cm,width=7.2cm]{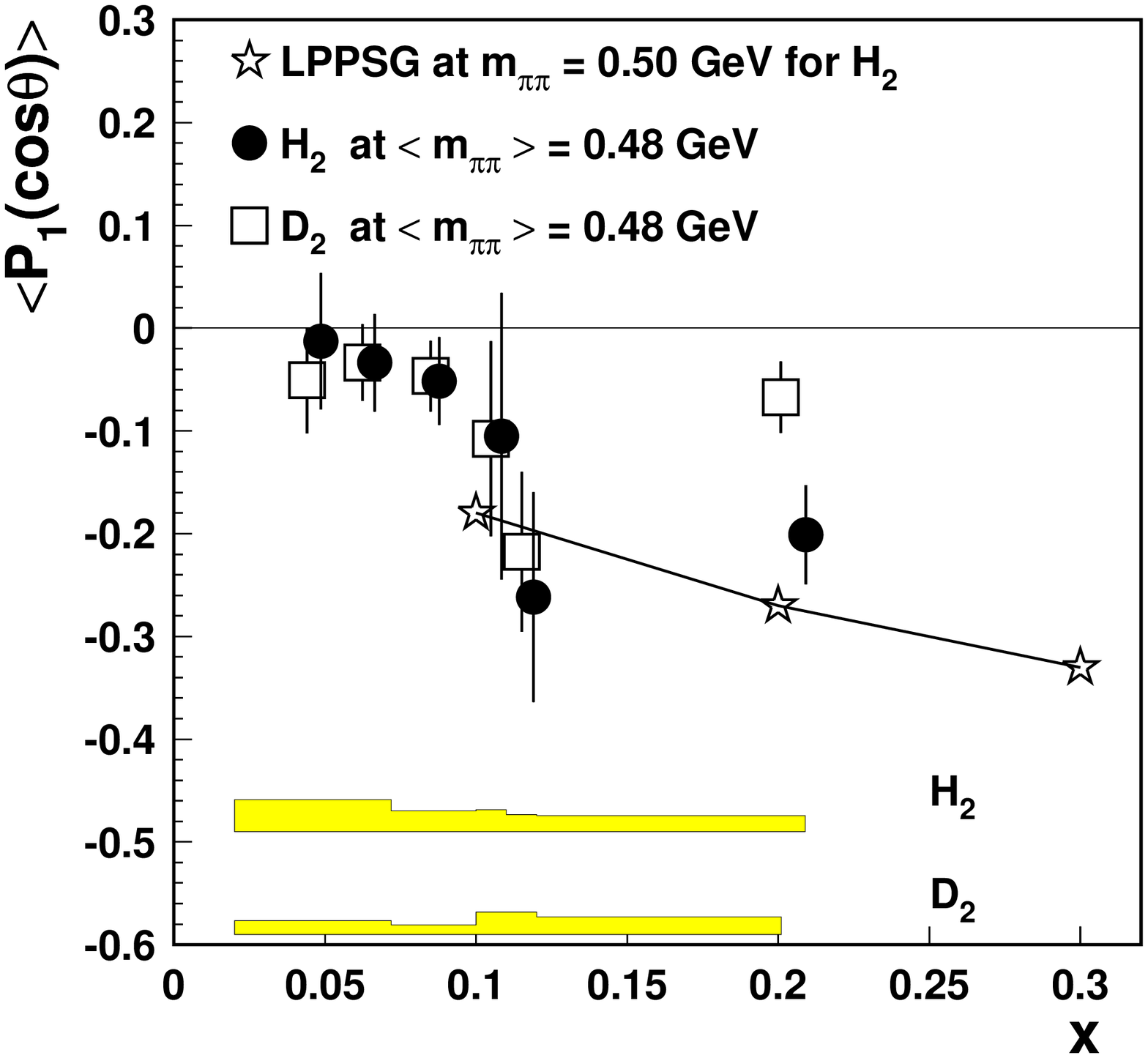}
\hspace{1.25cm}
  \includegraphics[height=7.8cm,width=7.2cm]{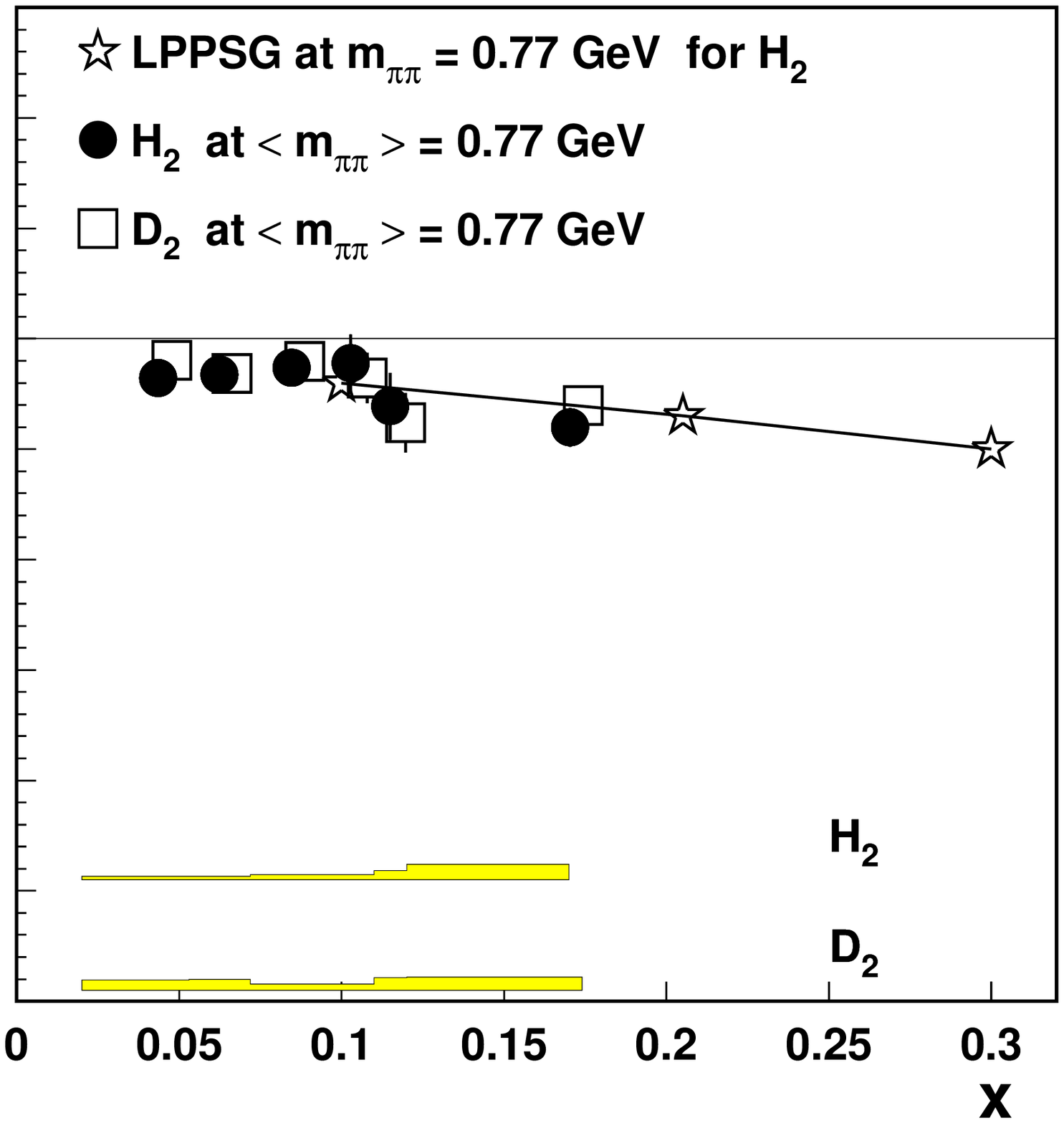}
  {\caption{The $x$-dependence of the Legendre moments
            $\langle P_1 \rangle$
            for both targets separately, in the regions
            $0.30 <m_{\pi\pi}<0.60$ GeV (left panel)
            and $0.60 <m_{\pi\pi}<0.95$ GeV (right panel).
            The systematic
            uncertainty is given by the error band.
            Theoretical predictions (stars) from
            LPPSG~\cite{polyakov00}
            for hydrogen, which neglect \mbox{two-gluon} exchange
            mechanism, are compared with the data.}
  \label{fig:xbj_results}}
 \end{center}
\end{figure*}

In order to study the contribution of the $f_2$ resonance
to the Legendre moments in more detail, 
the $m_{\pi\pi}$-dependence of the purely longitudinal 
combination \mbox{$\la P_1 + \frac{7}{3} \cdot P_3 \ra$} 
is presented in Fig.~\ref{fig:mpp_results_combinations} 
for both hydrogen and deuterium. 
For comparison, this figure also shows the combination
\mbox{$\la P_1 - \frac{14}{9} \cdot P_3 \ra$} which is believed 
to be dominated by the \mbox{higher-twist} transverse 
contribution to the excitation of the $f_2$ resonance. 
The comparison between these two distributions 
suggests that the  \mbox{higher-twist} transverse contribution 
to the Legendre moments in the $f_2$(1270) region 
is possibly as large as the longitudinal 
\mbox{leading-twist} production.
%

%
The $x$-dependence of $\langle P_1 \rangle$
is shown in Fig.~\ref{fig:xbj_results}
for both targets in two regions of $m_{\pi\pi}$:
\mbox{$0.30 <m_{\pi\pi}<0.60$ GeV} and
\mbox{$0.60 <m_{\pi\pi}<0.95$ GeV}.
The statistical precision at larger values of
$m_{\pi\pi}$ is insufficient for such a presentation.
The average values of $Q^2$, $-t$, and $x$
for both targets in these $m_{\pi\pi}$ regions
are reported in Table~\ref{tab_values}.
In both invariant mass regions and for both 
targets, $\la P_1 \ra$ is  non-zero, which we 
interpret as originating from the interference 
of resonant $\rho^0$ \mbox{$P$-wave} with 
non-resonant \mbox{$S$-wave} $\pi^+\pi^-$ 
production.
The moment increases in magnitude with $x$,  
suggesting that the exchange of flavour  
non-singlet quark combinations ($C=-1$) 
becomes competitive with the  dominant
singlet exchange ($C=+1$).
Predictions with only the quark exchange mechanism
in the GPD framework~\cite{polyakov00} 
are compared with the data, and are found to be 
in fair agreement with them. 
%

In summary, the Legendre moments 
$\langle P_{1}(\cos\theta) \rangle$ and   
$\langle P_{3}(\cos\theta) \rangle$ 
for exclusive electroproduction of 
$\pi^+\pi^-$ pairs have been measured for the 
first time for hydrogen and deuterium targets. 
The data show signatures of the interference 
between the dominant isospin state $I=1$ 
\mbox{($P$-wave)} and $I=0$ \mbox{($S,D$-wave)} 
of these pion pairs. 
The interference of the $\rho^0$ amplitude  
with the non-resonant \mbox{$S$-wave} 
and resonant \mbox{$D$-wave} states 
appears to be larger than the interference 
with the resonant $f_0$ \mbox{$S$-wave}. 
In the $f_2$ region, the combinations 
\mbox{$\la P_1 + 7/3 \cdot P_3 \ra$}
and
\mbox{$\la P_1 - 14/9 \cdot P_3 \ra$}
are sensitive to the longitudinal and 
the transverse states of a 
\mbox{$D$-wave} $\pp$ pair, respectively.  
Comparison of these combinations suggests 
that, at $\la Q^2 \ra = 3 $ GeV$^2$, 
the higher-twist transverse contribution  
to the Legendre moments in the $f_2$ domain 
can be as large as 
the leading-twist longitudinal contribution. 

These results constrain models for 
Generalized Parton Distributions, and may allow, by  
comparing the data with a larger statistical significance 
with the more accurate \mbox{next-to-leading} order 
predictions with and without 
the inclusion of the two-gluon mechanism, the separation 
of the contributions of two-gluon and $q\bar{q}$ exchange 
mechanisms, which are connected to the quark and gluon 
content of the nucleon.

\begin{table*}[hbt!]
  \begin{tabular}{|c|c|c|c|}
     \hline
     \multicolumn{4}{|c|}{{\bf $m_{\pi\pi}$-dependence analysis}}\\
     \hline 
     & & & \\
     Target & $\la Q^2 \ra$ [GeV$^2$] & $\la -t \ra$ [GeV$^2$] & 
     $\la x \ra$ \\
     \hline
     $H$ & $3.2$ & $0.43$ & $\ 0.16 \ $\\
     $D$ & $3.3$ & $0.29$ & $\ 0.16 \ $\\
     \hline
  \end{tabular}
 
  \vspace{0.5cm}
  \begin{tabular}{|c|c|c|c|c|}
     \hline
     \multicolumn{5}{|c|}{{\bf $x$-dependence analysis}} \\
     \hline
     & \multicolumn{2}{|c|}{} & \multicolumn{2}{|c|}{} \\
     & \multicolumn{2}{|c|}{\bf$\la m_{\pi\pi} \ra = 0.48$ [GeV]} &
       \multicolumn{2}{|c|}{\bf$\la m_{\pi\pi} \ra = 0.77$ [GeV]} \\
     & \multicolumn{2}{|c|}{} & \multicolumn{2}{|c|}{} \\ 
     \hline
     & & & & \\
     Target & $\la Q^2 \ra$ [GeV$^2$] & $\la -t \ra$ [GeV$^2$] 
            & $\la Q^2 \ra$ [GeV$^2$] & $\la -t \ra$ [GeV$^2$] \\
     \hline
     $H$ & $2.3$ & $0.42$ & $2.1$ & $0.27$ \\
     $D$ & $2.3$ & $0.39$ & $2.1$ & $0.22$ \\
     \hline
  \end{tabular}
  {\caption{Average values for $\la Q^2 \ra$, $\la -t \ra$, 
            and $\la x \ra$ measured in the $m_{\pi\pi}$- (upper table) 
            and $x$- (bottom table) dependence of Legendre moments 
            for hydrogen and deuterium targets.}
  \label{tab_values}
  }
  \vspace{0.5cm}
\end{table*}

We are grateful for inspiring and helpful discussions with
\mbox{M. Diehl}, \mbox{B. Lehmann-Dronke}, 
\mbox{B. Pire} and \mbox{M. V. Polyakov}. 
We gratefully acknowledge the DESY management for its 
support, the staff at DESY,  and the collaborating institutions 
for their significant effort.
This work was supported by the FWO-Flanders, Belgium; the Natural 
Sciences and Engineering Research Council of Canada; the ESOP, 
INTAS and TMR network contributions from the European Union; 
the German Bundesministerium f\"ur Bildung und Forschung; 
the Italian Instituto Nazionale di Fisica Nucleare (INFN);
Monbusho International Scientific Research Program, JSPS and 
Toray Science Foundation of Japan; 
the Dutch Foundation for Fundamenteel Onderzoek der Materie (FOM);
the U.K. Particle Physics and Astronomy Research Council; 
and the U.S. Department of Energy and National Science Foundation.
%


\end{document}